\documentclass[11pt]{article}

\usepackage[margin=1.25in]{geometry}
\usepackage{amsmath,amssymb,amsthm,mathtools}
\usepackage{enumitem}
\usepackage{xcolor}
\usepackage[authoryear,round]{natbib}
\usepackage{booktabs}
\usepackage{lmodern}
\usepackage[T1]{fontenc}

\usepackage{microtype}
\usepackage[onehalfspacing]{setspace}
\usepackage{tikz}
\usepackage{pgfplots}
\pgfplotsset{compat=1.18}
\usetikzlibrary{positioning,arrows.meta}
\usepackage[colorlinks=true,linkcolor=blue!55!black,citecolor=blue!55!black,urlcolor=blue!55!black]{hyperref}
\hypersetup{pdftitle={Coarse Price Discrimination},pdfauthor={Yaonan Jin and Yingkai Li}}

\newcommand{\Rstar}{R^{*}}
\newcommand{\LPP}{\mathrm{LP}}
\newcommand{\eps}{\varepsilon}
\newcommand{\E}{\mathbb{E}}
\renewcommand{\Pr}{\mathbb{P}}
\newcommand{\ind}[1]{\mathbf{1}\{#1\}}
\newcommand{\cF}{\mathcal{F}}

\newtheoremstyle{econ}%
  {\topsep}{\topsep}%
  {\itshape}{}%
  {\scshape}{:}%
  {.5em}{}
\theoremstyle{econ}
\newtheorem{theorem}{Theorem}
\newtheorem{lemma}{Lemma}

\newtheorem{proposition}{Proposition}
\newtheoremstyle{econdef}%
  {\topsep}{\topsep}%
  {\normalfont}{}%
  {\scshape}{:}%
  {.5em}{}
\theoremstyle{econdef}
\newtheorem{definition}{Definition}
\newtheorem{remark}{Remark}
\newtheorem{example}{Example}

\title{Coarse Price Discrimination\thanks{We are grateful to Yiding Feng, Jiashuo Jiang, and Wenjia Wang for invaluable discussions. Yaonan Jin acknowledges financial support from the HKUST Start-up Grant, and Yingkai Li acknowledges financial support from the NUS Start-up Grant. This paper was developed with the assistance of ChatGPT.}}
\author{
Yaonan Jin\thanks{Hong Kong University of Science and Technology. Email: {\tt jinyaonan@ust.hk}}
\and
Yingkai Li\thanks{National University of Singapore. Email: {\tt yk.li@nus.edu.sg}}
}
\date{\today}

\begin{document}
\maketitle

\begin{abstract}
\noindent We study the revenue cost of limiting a seller to $L$ posted prices when heterogeneous buyers with independent values compete for scarce inventory. Our large-market assumption requires that removing any buyer have little effect on optimal revenue. We first make buyers arbitrarily replaceable at each fixed stock and then let stock grow. In this limit, one price guarantees half of optimal mechanism revenue and two guarantee $e/(e+1)$, approximately $73.1\%$. For $L\ge3$, the corresponding guarantee is a $1-1/(2L-1)$ fraction. The guarantees accommodate arbitrary differences in demand across buyers and do not require a separate price for each customer group.

\vspace{5pt}

\end{abstract}

\onehalfspacing

\section{Introduction}\label{sec:intro}

Differences in demand can make individually tailored prices valuable. A seller who can distinguish many customer groups may therefore wish to set a different price for each. We ask how closely a short price list can approach the revenue from unrestricted personalized offers when heterogeneous buyers compete for scarce inventory.

Firms often know more about their customers than their price lists reveal. \citet{DellaVignaGentzkow19} document nearly uniform prices within U.S. retail chains despite substantial differences across stores. Sephora's Savings Event offers three membership-tier discounts even though the firm collects detailed purchase histories and customer preferences.\footnote{The advertised member-tier discounts are 10\%, 15\%, and 20\%; product-specific and referral offers also exist. See Sephora's \href{https://www.sephora.com/beauty/savings-event-faqs}{Savings Event FAQs} and \href{https://www.sephora.com/beauty/privacy-policy}{Privacy Policy} (accessed October 31, 2026).} Maintaining more prices can require additional administration and communication and make differences in treatment harder to explain. The revenue cost of a short price list is especially difficult to assess when buyers draw on a common stock: a discount that attracts one customer can use inventory that would otherwise be available to another.

We study a seller with a fixed stock of identical goods and buyers who each want one unit. Values are private and drawn independently from known distributions that may differ arbitrarily across buyers. The seller chooses at most $L$ posted prices and assigns one offer to each buyer before values are observed. She offers the higher prices first and stops when stock is exhausted. Unrestricted personalized posted pricing allows a different price for every buyer under the same allocation rule, providing a direct benchmark for the cost of limiting price differentiation. Our guarantees hold against the stronger benchmark of the optimal incentive-compatible mechanism, which can also condition allocation and payments on buyers' reported values.

Our large-market assumption requires that losing any one buyer have little effect on optimal revenue. The condition permits a buyer to be valuable in isolation, provided other customers can replace her contribution. Consider, for example, any fixed collection of customer groups. Buyers within each group draw their values independently from the same distribution, while the distributions may differ arbitrarily across groups. As the number of buyers in every group grows, each individual becomes replaceable while the differences in demand across groups persist. Replication illustrates how the assumption can arise; the theorem does not require this group structure.

Our main result bounds the revenue loss from using $L$ prices uniformly over buyers' demand distributions. We first make buyers arbitrarily replaceable at each fixed stock and then let stock grow. For $L\ge3$, the resulting worst-case ratio of optimal revenue to the best $L$-price revenue lies between $1+1/(2L-1)$ and $1+1/(2L-2)$. Halving a small worst-case revenue shortfall therefore requires approximately twice as many prices. Explicit guarantees in the same limiting comparison show that three prices capture at least 80\% of optimal revenue and ten capture at least 94.7\%. In this asymptotic comparison, the number of prices needed for a given revenue guarantee need not grow with the number of customer groups.

The two benchmarks yield the same sharp asymptotic ratios (Theorem~\ref{thm:sharp-market}). The limiting loss therefore measures the cost of restricting price differentiation itself. For each fixed finite $L$, the sharpness construction gives markets in which even the best $L$ prices leave a nonvanishing revenue loss in this limit. The large-market assumption therefore does not eliminate the economic cost of restricting prices.

The main intuition concerns how much inventory is needed to collect a given amount of spending. We isolate it in a simple auxiliary market: buyers have known values per unit and budgets, and can purchase any quantity their budgets allow. At an acceptable price, a buyer spends her budget if enough stock remains. Lowering her price then preserves spending but increases the quantity needed to collect it. Grouping buyers at a common lower price has an inventory cost, even in a market with no uncertainty. An extra price can reduce this cost by preserving a higher price for some of the buyers who would otherwise be pooled.

The seller need not preserve every source of spending. When each buyer pays her value per unit, the lowest-value buyers use the most inventory per dollar spent. Forgoing some of that spending can release enough stock to pool the rest at a few prices. Balancing this cost of pooling against the inventory saved by excluding some demand yields the revenue guarantees. We first characterize this trade-off in the auxiliary market.

In the limiting comparison, this auxiliary trade-off also determines the worst-case revenue loss from coarse price discrimination in the original market. The connection works in both directions. Competition limits the revenue lost when auxiliary prices are implemented, and a separate comparison handles uncertain sales. Conversely, markets with private values can reproduce the auxiliary trade-off.

The finite-market bound separates inventory risk from the effect of competition (Proposition~\ref{prop:finite-bound}). Larger stock reduces the relative loss from uncertain purchases, while greater individual replaceability reduces the error from implementing auxiliary prices. When stock grows, competition must strengthen relative to inventory. Replicated customer groups provide one sufficient condition: every group grows faster than stock. Without such control, a fixed price list can capture a vanishing share of revenue even as individual removal losses become small (Remark~\ref{rem:nested}).

\subsection{Related literature}\label{sec:related}

The closest economic literature studies the relative profitability of uniform pricing and third-degree price discrimination. \citet{MaluegSnyder06} bound the gains from discrimination in terms of the number of markets or distinct discriminatory prices, under conditions on demand and costs. \citet{BergemannCastroWeintraub22} obtain a one-half profit guarantee for uniform pricing, independent of the number of segments, under common support and concavity of segment profit in the price. Our large-market assumption provides a different basis for a guarantee in a market with scarce inventory. The assumption accommodates heterogeneous supports, and our bounds quantify the revenue cost of an intermediate number of prices between uniform pricing and unrestricted discrimination.

Other work studies the information that makes discrimination possible. \citet{BergemannBM15} characterize the consumer and producer surplus combinations achievable by varying market segmentation while holding aggregate demand fixed. \citet{XieZhuShishkin25} examine pricing when the relationship between customer characteristics and values must be learned from samples, showing how estimation costs can make discrimination less profitable than uniform pricing with limited data. Our seller knows the demand distributions and chooses how to assign a limited set of prices across buyers. The restriction concerns the prices she implements, while allowing her to use all the information available about demand. This separates the revenue cost of coarse pricing from the design or acquisition of information.

The auction literature makes clear why a substantial guarantee requires more than a large stock. For a single item with independent values, \citet{AlaeiHNPY19} compare anonymous pricing with an ex-ante relaxation, and \citet{JinLQTX19} establish its tight revenue gap relative to the optimal auction under the distributional conditions considered there. For multiple units, \citet{JinJiangLuZhang22} show that the advantage of discrimination can grow without bound as stock increases, even within a distributional class for which unrestricted sequential posted pricing becomes asymptotically optimal. Our results identify a competitive regime in which the loss from one price remains bounded as stock grows and quantify the benefit of additional price levels. Our main guarantees use the stronger benchmark of the optimal mechanism characterized by \citet{Myerson81}, whose relation to monopoly pricing is developed by \citet{BulowRoberts89}.

\citet{BlumrosenHolenstein08} compare the asymptotic revenues of uniform and discriminatory posted prices with optimal single-item auctions for identically distributed buyers under extreme-value conditions. \citet{CorreaLivanosPizarroVerdugo25} study fixed-price welfare guarantees in large i.i.d.\ markets and derive revenue guarantees under suitable conditions on virtual values. These papers hold the common demand distribution fixed as the market grows. Our bounds apply uniformly over heterogeneous markets satisfying the large-market assumption, allowing distributions to vary across markets and quantifying the cost of restricting the number of prices.

Our notion of a large market is related to the diminishing revenue returns to adding buyers in matroid environments established by \citet{DughmiRoughgardenSundararajan12}. Their revenue-submodularity result helps explain why replication within customer groups makes each buyer's departure less consequential. Classical large-auction results, such as \citet{Swinkels01}, study convergence to allocative efficiency. Here the economic question is how closely a limited set of prices can approximate optimal revenue once buyers become individually replaceable. Our sharp bounds quantify how the remaining loss from coarse pricing depends on the number of prices.

Finally, the restriction on prices differs from the menu-size complexity studied by \citet{HartNisan19}, where a single buyer with multidimensional values chooses among allocation-payment options. In our setting, each buyer receives one assigned offer, and simplicity concerns the number of prices used across buyers. The analysis builds on the sequential posted-pricing framework of \citet{ChawlaHMS10} and the methods of \citet{Alaei14,Yan11} for relating expected allocations to feasible sales under uncertain demand.

\section{Model}\label{sec:model}

A seller has a \emph{stock} of $k\ge1$ identical units of an indivisible good and faces $n\ge1$ buyers, each demanding one unit. Buyer $i$ privately knows her value $v_i\ge0$, drawn from a distribution $F_i$ with finite mean. Buyers are risk neutral and have quasilinear utility.

The seller observes each buyer's identity and knows the distribution profile $F=(F_1,\ldots,F_n)$. Distribution $F_i$ summarizes her information about buyer $i$, including observable characteristics or purchase history. Conditional on this information, values are independent across buyers, and their distributions may differ arbitrarily. Buyers with the same characteristics may share a distribution while having different private values.

\subsection{Price discrimination}\label{sec:mechanisms}

Buyers receive enforceable, buyer-specific offers. This describes settings in which the seller can verify identity or eligibility and prevent buyers from switching to another customer's price. Each buyer can accept or reject her assigned offer, and rejection does not give access to another price.

\paragraph{Unrestricted price discrimination.} The seller fixes a price vector $p=(p_1,\dots,p_n)\in[0,\infty]^n$ before values are observed, allowing a different price for every buyer. She approaches buyers in weakly decreasing order of their assigned prices, breaking ties by index and skipping buyers with $p_i=\infty$. An approached buyer accepts if and only if $v_i\ge p_i$, and the seller stops after $k$ sales. Write $\LPP_{\infty}(F)$ for the supremum expected revenue over all such price vectors.

Control over service priority is part of the environment. The seller may, for example, collect binding purchase requests before allocating stock or choose which customer groups receive offers first. Such arrangements allow higher-price offers to receive priority even when buyers arrive at different times.

\paragraph{Coarse price discrimination.} The seller chooses a limited number of price levels and assigns buyers to them using the full profile $F$. Buyers with different demand distributions may therefore share a price; price categories need not coincide with pre-existing customer groups.
\begin{definition}[$L$-level mechanism]\label{def:LP}
For $L\ge1$, an \emph{$L$-level mechanism} is a sequential posted-price mechanism of the form above whose price vector has at most $L$ distinct finite entries.
\end{definition}
We write $\LPP_L(F)$ for the supremum expected revenue over such price vectors. The value $p_i=\infty$ represents no offer and does not count as a price level. With one level, all buyers who receive an offer face the same price; allowing up to $n$ levels permits a different price for every buyer and gives $\LPP_n(F)=\LPP_{\infty}(F)$.

\paragraph{Revenue benchmarks.} Unrestricted personalized posted pricing, $\LPP_{\infty}(F)$, provides the direct benchmark for restricting the number of prices. We also compare coarse pricing with optimal revenue from incentive-compatible mechanisms, which can condition allocations and payments on buyers' reported values.

By the revelation principle \citep{Myerson81}, we may restrict attention to direct mechanisms $(X,P)$. At each report profile $\hat v\in[0,\infty)^n$, $X_i(\hat v)\in[0,1]$ and $P_i(\hat v)\in\mathbb R$ are buyer $i$'s allocation probability and expected payment, averaging over internal randomization. Feasibility requires at most $k$ buyers to receive a unit in every realization.

Given truthful reports by the other buyers, buyer $i$'s expected utility is
\begin{equation*}
u_i(v_i,\hat v_i):=\E_{v_{-i}\sim F_{-i}}\bigl[v_iX_i(\hat v_i,v_{-i})-P_i(\hat v_i,v_{-i})\bigr].
\end{equation*}
The mechanism is \emph{Bayesian incentive compatible} and \emph{interim individually rational} if, for every $i$ and all $v_i,\hat v_i\ge0$, $u_i(v_i,v_i)\ge u_i(v_i,\hat v_i)$ and $u_i(v_i,v_i)\ge0$. The \emph{optimal revenue} is
\begin{equation*}
\Rstar(F):=\sup_{(X,P)}\E_{v\sim F}\left[\sum_{i=1}^n P_i(v)\right],
\end{equation*}
where the supremum is over feasible direct mechanisms satisfying these two conditions.

\subsection{Large-market assumption}\label{sec:large-market}

We measure competition by the revenue the seller would lose if any one buyer left. The seller may redesign her selling mechanism after that departure, keeping the same stock and the same information about the remaining buyers. Our large-market assumption requires that their demand can sustain nearly the same optimal revenue.
\begin{definition}[Large-market assumption]\label{def:large-market}
For $\eps\ge0$, a market $F$ is \emph{$\eps$-large} if
\begin{equation*}
\Rstar(F_{-i})\;\ge\;(1-\eps)\,\Rstar(F)\qquad\text{for every buyer }i.
\end{equation*}
\end{definition}
Equivalently, the removal loss $\Delta_i(F):=\Rstar(F)-\Rstar(F_{-i})$ is at most $\eps\Rstar(F)$. A smaller $\eps$ means that each buyer's contribution is more readily replaced. This permits substantial revenue from a buyer when served alone. For example, suppose $k+1$ buyers have deterministic value $a>0$. Optimal revenue is $ka$ both before and after any buyer leaves, although each buyer alone can generate revenue $a$. This market is $0$-large.

Replication provides another way for the assumption to arise. Consider a fixed collection of customer groups, with buyers within each group drawing values independently from a common distribution. Expanding every group gives the seller more substitutes for each customer, making an individual departure less consequential relative to optimal revenue at a fixed stock. Demand distributions can remain arbitrarily different across groups, so the differences that motivate personalized pricing persist. Section~\ref{sec:joint-growth} provides the formal calculation. The assumption itself requires neither a group structure nor identical distributions within groups.

\section{The Inventory Cost of Coarse Pricing}\label{sec:auxiliary}

To isolate the inventory cost of pooling buyers at common prices, we introduce an auxiliary market with known values and spending budgets. Lowering an acceptable price preserves the budget a buyer is willing to spend but increases the inventory required to collect it. The seller can accommodate this extra demand by forgoing some spending from lower-value buyers. The auxiliary budgets will represent revenue opportunities when we return to the original market. The original buyers still have private values and unit demand, with no budget constraint.

\subsection{The auxiliary environment}\label{sec:auxiliary-model}

An instance $I=((w_i,\beta_i)_{i=1}^n,K)$ consists of an allocation capacity $K>0$ and finitely many buyers. Allocations may be fractional, with total quantity at most $K$. Buyer $i$ has a deterministic per-unit value $w_i>0$ and a deterministic budget $\beta_i>0$, both known to the seller. Her value for quantity $q\ge0$ is $w_iq$, and she has no quantity limit apart from her budget. At an assigned price $p>0$, her desired quantity is $\beta_i/p$ if $p\le w_i$ and zero otherwise.

Thus a buyer spends her full budget whenever the price is acceptable and sufficient inventory remains. We adopt purchase at equality $p=w_i$. If inventory is insufficient, she buys the remaining quantity and pays the assigned price per unit.

The seller assigns one price to each buyer, with $p_i=\infty$ denoting no offer, and serves buyers in decreasing assigned-price order, breaking ties by index. At most $L$ distinct finite prices may be used. Write $\LPP^{X}_L(I)$ for the supremum revenue of these policies and $\LPP^{X}_{\infty}(I)$ for the unrestricted-price benchmark, with the superscript $X$ indicating the auxiliary market. Revenue in the auxiliary market is deterministic.

The stock restriction is essential: without it, a single positive price below every buyer's value would collect all budgets. With scarce inventory, an optimal policy with unrestricted auxiliary prices charges each buyer her value and serves buyers in decreasing value order, collecting budgets until stock runs out. Lemma~\ref{lem:auxiliary-full} formalizes this observation. Capacity suffices to collect every budget at its buyer's value when
\begin{equation}\label{eq:auxiliary-affordable}
\sum_{i=1}^n\frac{\beta_i}{w_i}\le K,
\end{equation}
so unrestricted revenue is the total spending available, $B:=\sum_{i=1}^n\beta_i$. This case makes the inventory cost of pooling particularly transparent; the guarantee below applies to every auxiliary instance, including those with less stock.

A dollar collected at price $p$ uses $1/p$ units. The following example shows how this inventory cost makes both the number and the placement of prices matter.

\begin{example}[What extra prices buy]\label{ex:whatLbuys}
There are three buyers, with values $6,4,2$ and budgets $6,8,6$, respectively. Charging each buyer her value uses quantities $1,2,3$, so total unrestricted revenue is $B=20$. Let stock be $K=7$.

The table assigns each buyer the highest listed price she accepts. The inventory requirement is the quantity needed to collect all eligible budgets; when it exceeds $K$, service stops once stock is exhausted.
\begin{center}
\begin{tabular}{@{}lccc@{}}
\toprule
number of prices & price list & inventory requirement & revenue\\
\midrule
three & $6,4,2$ & $1+2+3=6$ & $20$\\
two & $4,2$ & $14/4+6/2=6.5$ & $20$\\
two & $6,2$ & $6/6+14/2=8$ & $18$\\
one & $4$ & $14/4=3.5$ & $14$\\
one & $2$ & $20/2=10$ & $14$\\
\bottomrule
\end{tabular}
\end{center}
The first buyer spends six dollars at any of these prices, but purchases more units at a lower price. Pooling her with the middle buyer at $4$ leaves enough stock to collect all budgets. Keeping her price at $6$ while pooling the other two buyers at $2$ instead exhausts stock at revenue $18$. Thus the placement of prices matters even when the lowest price is unchanged.

One price earns at most $14$: prices above $4$ can collect only the first budget, prices in $(2,4]$ can collect at most the first two budgets, and prices at most $2$ earn at most $2K=14$. Price $4$ achieves this bound by excluding the lowest-value buyer; price $2$ achieves it by selling all stock. Two well-chosen prices collect all $20$. A third price leaves revenue unchanged but lowers the inventory requirement from $6.5$ to $6$.
\end{example}

\subsection{Revenue guarantees and sharp constants}\label{sec:auxiliary-guarantee}

The next proposition characterizes the worst-case revenue cost of restricting the price list. Its bounds do not depend on the number of buyers or on how their values and budgets differ.

\begin{proposition}[The sharp cost of coarse prices]\label{thm:auxiliary}
For each $L\ge1$, define the worst-case revenue ratio
\begin{equation*}
C_L^\star:=\sup_I
\frac{\LPP^{X}_{\infty}(I)}{\LPP^{X}_L(I)},
\end{equation*}
where the supremum is over nonempty finite auxiliary instances with positive stock, values, and budgets. Then
\[
C_1^\star=2,\qquad C_2^\star=1+\frac1e.
\]
For $L\ge3$,
\begin{equation*}
1+\frac1{2L-1}\le C_L^\star\le1+\frac1{2L-2}.
\end{equation*}
\end{proposition}

Thus one price guarantees half of unrestricted revenue, and two guarantee the fraction $e/(e+1)$. These two guarantees are sharp. For $L\ge3$, the exact constants remain open; the stated lower bound need not be exact either, as Proposition~\ref{prop:det-tight} shows. The bounds imply that halving a small worst-case revenue shortfall requires approximately twice as many prices.

\subsection{Pooling, exclusion, and inventory}\label{sec:apply-auxiliary}

\begin{figure}[t]
\centering
\begin{tikzpicture}[x=1.05cm,y=1.05cm,>=Stealth,font=\small]
\fill[blue!16] (0,0) rectangle (1,.65);
\fill[blue!16] (1,0) rectangle (2,.95);
\fill[blue!16] (2,0) rectangle (3,1.35);
\fill[blue!16] (3,0) rectangle (4,1.70);
\fill[blue!16] (4,0) rectangle (5,2.15);
\fill[blue!16] (5,0) rectangle (6,2.60);
\fill[orange!38] (0,.65) rectangle (1,1.35);
\fill[orange!38] (1,.95) rectangle (2,1.35);
\fill[orange!38] (3,1.70) rectangle (4,2.60);
\fill[orange!38] (4,2.15) rectangle (5,2.60);
\fill[green!22!white] (6,0) rectangle (7,3.00);
\fill[green!22!white] (7,0) rectangle (8,3.50);
\draw[->] (-.15,0)--(8.45,0) node[right] {budget rank};
\draw[->] (0,-.1)--(0,3.85);
\node[rotate=90] at (-1,1.9) {units per dollar};
\draw[very thick,blue!65!black]
 (0,.65)--(1,.65)--(1,.95)--(2,.95)--(2,1.35)--(3,1.35)
 --(3,1.70)--(4,1.70)--(4,2.15)--(5,2.15)--(5,2.60)--(6,2.60)
 --(6,3.00)--(7,3.00)--(7,3.50)--(8,3.50);
\draw[thick,orange!85!black] (0,1.35)--(3,1.35);
\draw[thick,orange!85!black] (3,2.60)--(6,2.60);
\draw[dashed,gray] (3,0)--(3,2.60);
\draw[dashed,gray] (6,0)--(6,3.00);
\draw[dashed,green!45!black] (6,2.60)--(8,2.60);
\node[left] at (0,2.60) {$u(b)$};
\node[below] at (0,0) {$0$};
\node[below] at (3,0) {$R_1$};
\node[below] at (6,0) {$b$};
\node[below] at (8,0) {$B$};
\draw[<->,gray] (0,-.60)--(6,-.60) node[midway,below] {selected spending};
\draw[<->,gray] (6,-.60)--(8,-.60) node[midway,below] {forgone spending};
\fill[blue!16] (.1,4.03) rectangle (.35,4.23);
\node[anchor=west] at (.43,4.13) {ideal inventory};
\fill[orange!38] (2.85,4.03) rectangle (3.10,4.23);
\node[anchor=west] at (3.18,4.13) {pooling cost};
\fill[green!22!white] (5.55,4.03) rectangle (5.80,4.23);
\node[anchor=west] at (5.88,4.13) {inventory released};
\end{tikzpicture}
\caption{The rising steps give inventory per dollar at individual values. The selected spending $(0,b]$ is pooled into blocks $(0,R_1]$ and $(R_1,b]$, each charged its endpoint price. Orange areas are the extra inventory required by pooling. The green area is the inventory released by forgoing $(b,B]$; it is at least $(B-b)u(b)$, the rectangle below the dashed green line.}
\label{fig:capacity-cost}
\end{figure}
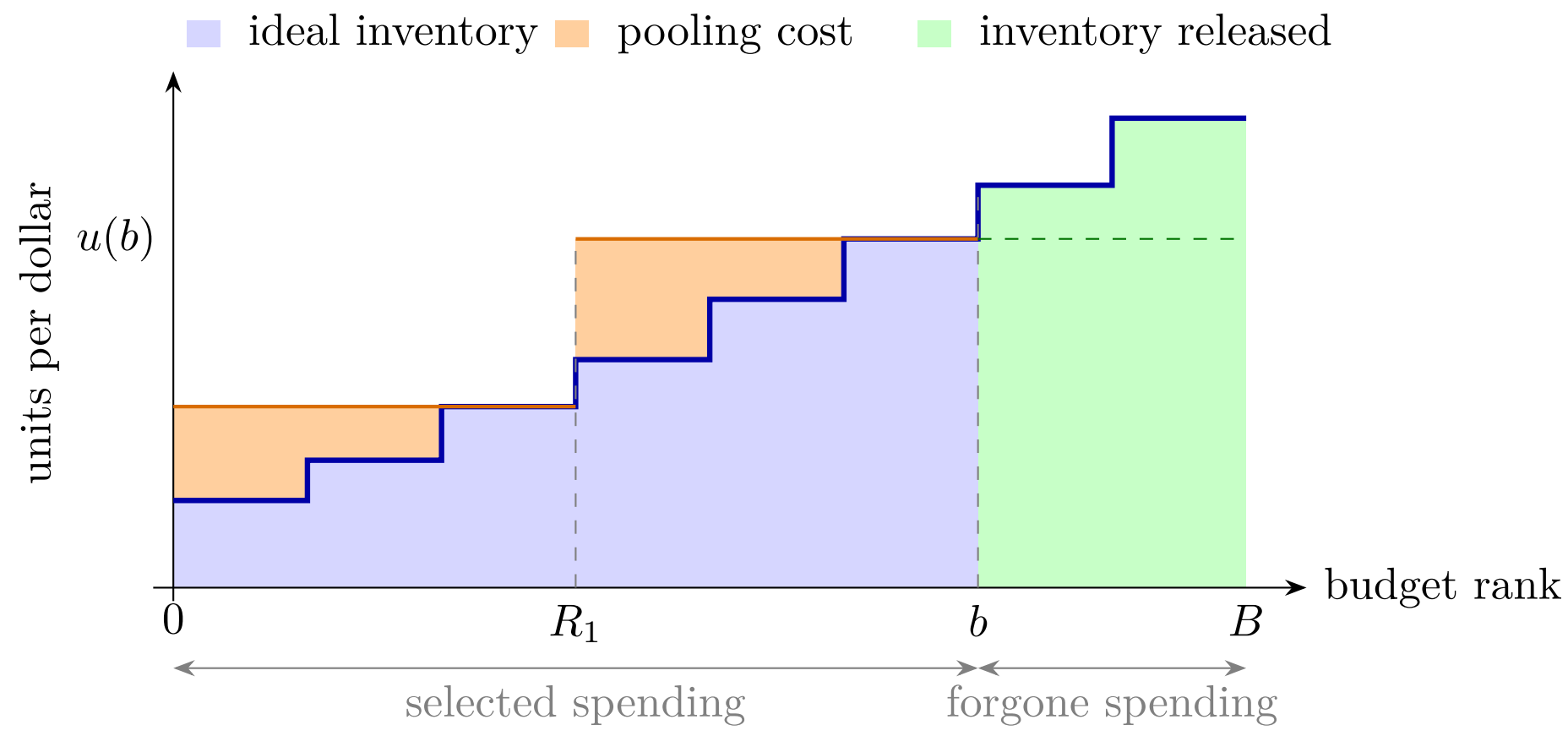

For any finite auxiliary instance satisfying~\eqref{eq:auxiliary-affordable}, write $B:=\sum_{i=1}^n\beta_i$ for its total budget. Arrange its buyers in decreasing value order, breaking ties by index, and place their budgets consecutively on $(0,B]$. Buyer $i$ occupies an interval of length $\beta_i$. At budget rank $R$, let $w(R)$ be the corresponding buyer's value. Collecting one dollar from her at that value uses $1/w(R)$ units of inventory, giving the \emph{unit-cost profile}
\begin{equation*}
u(R):=\frac1{w(R)},\qquad 0<R\le B,
\qquad u(0):=u(0^+).
\end{equation*}
Each budget interval has the form $(z,y]$, so an upward jump takes effect immediately to its right. The profile is nondecreasing and left-continuous. Its area is the quantity needed to collect all budgets at individual values:
\begin{equation*}
\int_0^B u(R)\,dR
=\sum_{i=1}^n\frac{\beta_i}{w_i}\le K.
\end{equation*}
Figure~\ref{fig:capacity-cost} illustrates a plan that collects the first $b$ budget dollars at two prices, with a boundary at $R_1$. Charging each block its lowest buyer value adds the orange inventory cost. Forgoing the final $B-b$ dollars releases the green inventory. The quantity used at individual values plus the pooling cost must fit within stock $K$.

\begin{definition}[Price plans]\label{def:plan}
A plan divides selected spending into at most $L$ groups, each charged a common price. It chooses $L$ boundary positions
\[
0=R_0\le R_1\le\cdots\le R_L=b\le B.
\]
Each nonempty block $(R_{\ell-1},R_\ell]$ uses price $p_\ell=1/u(R_\ell)$. Its buyers accept this price because their values are at least the value at its endpoint. The plan's revenue is $b$, and its inventory requirement is
\[
N:=\sum_{\ell=1}^L u(R_\ell)(R_\ell-R_{\ell-1}).
\]
It is feasible if $N\le K$. Repeated boundaries represent empty blocks, so $L$ positions allow at most $L$ distinct prices. The maximum spending collectible with at most $L$ prices and capacity $K$ is
\begin{equation*}
V_L=V_L(u,K):=\max\{b:N\le K\}.
\end{equation*}
A boundary may split a buyer's budget into portions. Empty blocks contribute zero and need no price evaluation.
\end{definition}

The maximum exists by Remark~\ref{rem:attainment}. The next lemma identifies the plan value with auxiliary revenue and implements budget portions with one assigned price per buyer.

\begin{lemma}[Canonical price plans]\label{lem:auxiliary-plans}
For a finite auxiliary instance satisfying~\eqref{eq:auxiliary-affordable},
\begin{equation*}
\LPP^{X}_L(I)=V_L(u,K).
\end{equation*}
Every feasible plan with revenue $b$ can be implemented by a policy earning at least $b$, with one assigned price per buyer and at most $L$ distinct prices.
\end{lemma}

The proof is in Appendix~\ref{app:covering}.

The inventory comparison in Figure~\ref{fig:capacity-cost} guides how much spending to forgo so that the released inventory covers the cost of pooling. We first prove the upper bound for $L\ge3$, then illustrate the loss from pooling with a simple lower-bound construction. Appendix~\ref{app:covering} supplies the exact one- and two-price calculations and the sharper finite-$L$ lower bound in Proposition~\ref{thm:auxiliary}.

\begin{proof}[Proof of the upper bound for $L\ge3$]
For the upper bound, restrict each budget to the payment collected under unrestricted auxiliary pricing, and let $B$ be the sum of those payments. The resulting budgets are affordable at individual values. Restoring the original budgets afterward cannot lower the revenue of a fixed price assignment. We may therefore work with a nondecreasing unit-cost profile $u$ satisfying $\int_0^B u(R)\,dR\le K$.

For $L\ge3$, we forgo enough of the lowest-value spending to release the inventory needed to pool the remaining buyers, as illustrated in Figure~\ref{fig:capacity-cost}. Write $b$ for the selected spending and $d$ for the spacing between internal price boundaries. Set
\[
b:=\frac{2L-2}{2L-1}B,\qquad d:=\frac{b}{L-1}.
\]
Place the $L-1$ internal price boundaries at $\xi,\xi+d,\ldots,\xi+(L-2)d$, where $\xi$ is uniform on $(0,d)$, and add endpoints $0$ and $b$. Charge each block the value at its last budget dollar. An upward jump in unit cost adds inventory equal to its height times the distance back to the preceding boundary. Shifting the grid uniformly makes that distance at most half a grid spacing on average, or $d/2$. This is the source of the coefficient $1/2$. Since the jump heights sum to at most $u(b)$, the expected extra inventory is at most $du(b)/2$. The choice of $b$ makes the spending forgone release at least this much inventory:
\[
\int_b^B u(R)\,dR\ge(B-b)u(b)=\frac d2u(b).
\]
Some fixed partition therefore fits within stock $K$ and collects $b$, proving $C_L^\star\le B/b=1+1/(2L-2)$. Lemma~\ref{lem:auxiliary-plans} implements the partition with one price per buyer, including when a boundary splits a budget.

\end{proof}

\paragraph{A lower-bound illustration.}
For $L\ge3$, the following construction gives $C_L^\star\ge1+1/(2L)$, a weaker bound than in Proposition~\ref{thm:auxiliary}. Let unit cost rise linearly over the first dollar of spending. Pooling an interval of width $h$ then creates a triangular extra inventory cost $h^2/2$. Equal block widths minimize the sum of these costs, giving $1/(2L)$ with $L$ prices. We append $1/(2L)$ dollars of spending at unit cost one: collecting this final spending requires exactly the extra inventory that pooling the first dollar would consume. Formally, consider the limiting unit-cost profile
\[
u(R)=\min\{R,1\}\quad(0<R\le B),
\qquad B:=1+\frac1{2L},\qquad K:=\frac12+\frac1{2L}.
\]
The profile has total spending $B$ and area $K$. Applying the same price-plan formula to this profile, dividing $(0,1]$ into $L$ equal blocks uses exactly $K$ units and collects one dollar.

Appendix~\ref{app:linear-profile} verifies that this plan is optimal for the profile, so $V_L(u,K)=1$. Finite positive step approximations give auxiliary markets with revenue ratios approaching $B$; Proposition~\ref{prop:auxiliary-sharp} justifies this passage. Hence $C_L^\star\ge1+1/(2L)$.

\section{Revenue Guarantees in Large Markets}\label{sec:results}

We return to buyers with private values and unit demand. Lowering a price can now reduce expected revenue. The large-market assumption controls the revenue lost when auxiliary prices are implemented, allowing us to derive revenue guarantees for coarse price discrimination in the original market. Section~\ref{sec:joint-growth} derives the theorem from a finite-market revenue bound and a matching sharpness statement. Sections~\ref{sec:argument} and~\ref{sec:coarseness-constants} prove these two ingredients.

To state the results, let $D_i(p):=\Pr_{v_i\sim F_i}(v_i\ge p)$ denote buyer $i$'s demand at price $p\ge0$, measured by her acceptance probability before applying the stock constraint. Her quantile price $v_i(q)$ is the highest price accepted with probability at least $q$. Multiplying that price by the service probability $q$ gives her revenue curve:
\[
v_i(q):=\sup\{p\ge0:D_i(p)\ge q\},
\qquad R_i(q):=q\,v_i(q),\qquad q\in(0,1],
\]
with $R_i(0):=0$.

\begin{definition}[Regularity]\label{def:regular}
$F_i$ is \emph{regular} if its revenue curve $R_i$ is concave on $[0,1]$.
\end{definition}

Regularity is a standard assumption in optimal auction theory \citep{Myerson81}. Economically, it imposes diminishing marginal revenue as the probability of sale increases, linking optimal auction design to monopoly pricing \citep{BulowRoberts89}.\footnote{Regularity serves a technical role in simplifying the argument and can be substantially relaxed. Appendix~\ref{app:quasireg} establishes the same revenue bounds and sharp limiting ratios for the broader class of quasi-regular distributions, which includes order statistics of independent regular draws.}

For a fixed stock $k$ and $\eps\ge0$, write $\cF_k(\eps)$ for the class of independent regular $\eps$-large markets with finite means and $\Rstar(F)>0$. The positive-revenue restriction makes the revenue ratios below well defined; zero-revenue markets satisfy the corresponding revenue guarantees trivially.

\subsection{The sharp cost of restricting prices}\label{sec:auxiliary-sharpness}\label{sec:reading}

Competition first becomes arbitrarily strong at each fixed stock, controlling the loss from implementing auxiliary prices. Stock then grows, removing the remaining inventory risk.

\begin{theorem}[The cost of coarse price discrimination]\label{thm:sharp-market}
For every fixed $L\ge1$,
\[
\lim_{k\to\infty}\ \lim_{\eps\downarrow0}\ \sup_{F\in\cF_k(\eps)}
\frac{\Rstar(F)}{\LPP_L(F)}
=
\lim_{k\to\infty}\ \lim_{\eps\downarrow0}\ \sup_{F\in\cF_k(\eps)}
\frac{\LPP_{\infty}(F)}{\LPP_L(F)}=C_L^\star.
\]
\end{theorem}

Both benchmarks yield the same sharp guaranteed revenue share $1/C_L^\star$, so the limiting loss measures the cost of restricting price differentiation itself. Competition limits the revenue lost when auxiliary prices are offered to the original buyers, and growing stock reduces the loss from uncertain purchases. The remaining loss is the auxiliary inventory cost of pooling demand. Since $C_L^\star>1$ for every finite $L$, this cost can persist even when buyers become arbitrarily replaceable. Section~\ref{sec:coarseness-constants} constructs markets approaching it despite allowing the seller to optimize both prices and their assignment.

Table~\ref{tab:constants} translates Proposition~\ref{thm:auxiliary} into revenue shares. The guarantees are sharp for one and two prices and are lower bounds on the sharp share for larger $L$.

\begin{table}[!ht]
\centering
\begin{tabular}{@{}ccc@{}}
\toprule
$L$ & ratio bound in the nested limit & guaranteed revenue share\\
\midrule
$1$ & $2$ & $50.0\%$\\[4pt]
$2$ & $1+\dfrac1e$ & $73.1\%$\\[4pt]
$3$ & $\dfrac54$ & $80.0\%$\\[4pt]
$4$ & $\dfrac76$ & $85.7\%$\\[4pt]
$5$ & $\dfrac98$ & $88.9\%$\\[4pt]
$10$ & $\dfrac{19}{18}$ & $94.7\%$\\
\bottomrule
\end{tabular}
\caption{Explicit revenue guarantees after the large-market limit and then the large-stock limit. The displayed shares equal $1/C_L^\star$ for $L=1,2$ and are lower bounds on that sharp share for $L\ge3$.}
\label{tab:constants}
\end{table}

The guarantees concern total benchmark revenue and are uniform over the permitted demand distributions. Buyer assignments may still use all available demand information.

\subsection{Inventory risk and competition}\label{sec:joint-growth}

At finite stock, the guarantee reflects both uncertain purchases and the seller's dependence on individual buyers. The next bound separates these effects from the auxiliary cost of a short price list.

\begin{proposition}[Finite-market revenue bound]\label{prop:finite-bound}
For integers $k,L\ge1$, let
\[
c_k:=1-\frac{k^k}{e^k k!},
\qquad \eps_0(k):=\{k[2(2+\sqrt2)]^3\}^{-1}.
\]
For every $0\le\eps\le\eps_0(k)$ and $F\in\cF_k(\eps)$,
\begin{equation*}
\frac{\Rstar(F)}{\LPP_L(F)}
\le\frac{C_L^\star}{c_k}
+8\left(2+\sqrt2\right)(k\eps)^{1/3}.
\end{equation*}
\end{proposition}

The factor $c_k$ accounts for inventory risk. Even when expected demand fits the available stock, realized purchases can exceed it, preventing the seller from collecting every potential payment. Larger inventories reduce the loss allowed by this bound: $c_k\to1$, so the revenue loss $1-c_k$ and the ratio penalty $1/c_k-1$ vanish. Lemma~\ref{lem:cg} gives the revenue comparison under uncertain purchases.

The additional term reflects individual replaceability. At a fixed stock, stronger competition provides substitutes for each buyer, reducing $\eps$ and the revenue shortfall from implementing the auxiliary prices. If stock also grows, competition must strengthen sufficiently to keep $k\eps$ small. In particular, $k\to\infty$ and $k\eps\to0$ make both corrections vanish, giving the same asymptotic upper bound $C_L^\star$. The cube-root rate comes from balancing the errors in our proof; neither its exponent nor its coefficient is claimed optimal. The bound also applies against unrestricted personalized posted pricing because $\LPP_{\infty}(F)\le\Rstar(F)$.

The finite-market bound supplies the upper bounds in Theorem~\ref{thm:sharp-market}. The reverse inequalities require markets whose loss from restricting prices approaches the auxiliary loss. The following proposition provides them; Section~\ref{sec:coarseness-constants} gives the construction.

\begin{proposition}[Sharpness in large markets]\label{prop:market-sharpness}
For every integer $L\ge1$ and $\eta>0$, there is an integer $k_0\ge1$ such that, for every integer $k\ge k_0$ and every $\eps>0$, some $F\in\cF_k(\eps)$ satisfies
\[
\frac{\LPP_\infty(F)}{\LPP_L(F)}>C_L^\star-\eta.
\]
\end{proposition}

\begin{proof}[Proof of Theorem~\ref{thm:sharp-market}]
Both inner limits exist because the classes $\cF_k(\eps)$ shrink as $\eps$ decreases. Taking the supremum in Proposition~\ref{prop:finite-bound} and then $\eps\downarrow0$ at fixed $k$ gives
\[
\lim_{\eps\downarrow0}\sup_{F\in\cF_k(\eps)}
\frac{\Rstar(F)}{\LPP_L(F)}\le\frac{C_L^\star}{c_k}.
\]
Proposition~\ref{prop:market-sharpness} gives a lower bound $C_L^\star-\eta$ for the inner limit of $\LPP_\infty(F)/\LPP_L(F)$ at every sufficiently large $k$. Since $\LPP_\infty(F)\le\Rstar(F)$, the inner limits of both ratios lie between these bounds. Let $k\to\infty$ and then $\eta\downarrow0$, using $c_k\to1$.
\end{proof}

The customer groups in Section~\ref{sec:large-market} illustrate how competition can reduce $\eps$. Suppose group $g$ has $n_g$ buyers with independent values drawn from a common distribution $H_g$, which may differ arbitrarily across groups. For a buyer $i$ in group $g$,
\begin{equation*}
\Delta_i(F)\le\frac{\Rstar(F)}{n_g}.
\end{equation*}
Indeed, symmetry makes removal losses equal within a group, while $\sum_{j=1}^n\Delta_j(F)\le\Rstar(F)$ by revenue submodularity \citep{DughmiRoughgardenSundararajan12}. Thus $\eps=1/\min_g n_g$ suffices, and expanding every group strengthens individual replaceability while preserving differences in demand. When stock grows, it is sufficient for the smallest group to grow faster than inventory, so that $k/\min_g n_g\to0$.

A large stock alone gives no uniform guarantee for a fixed number of prices. The following example also shows why individual replaceability must be considered relative to inventory.

\begin{remark}[Competition relative to stock]\label{rem:nested}
Take $k$ units and $k$ buyers with deterministic values $v_i=1/i$. Their distributions are regular and optimal revenue is $H_k:=\sum_{i=1}^k1/i$, which grows on the order of $\log k$. Removing buyer $i$ costs $1/i\le1$, so the market is $\eps_k$-large for $\eps_k=1/H_k\to0$. At any positive posted price $p$, at most $\min\{k,\lfloor1/p\rfloor\}$ buyers accept, giving revenue at most
\[
p\min\{k,\lfloor1/p\rfloor\}\le1.
\]
Thus an $L$-level mechanism earns at most $L$, and its share of optimal revenue vanishes for fixed $L$. Inventory risk is absent here: every buyer can be served and values are known. The loss comes from the growing number of value scales that a short price list must cover. Although $\eps_k\to0$, the product $k\eps_k$ diverges. Small individual removal losses alone therefore do not ensure the guarantee along arbitrary joint growth of buyers and stock.
\end{remark}

\section{Proof of the Finite-Market Bound}\label{sec:argument}\label{sec:outline}\label{sec:proof-setup}

To prove Proposition~\ref{prop:finite-bound}, we need an auxiliary market whose unrestricted revenue nearly matches optimal mechanism revenue and whose coarse prices can be offered to the original buyers with little loss. These two properties let us apply the inventory comparison from Section~\ref{sec:auxiliary}. We first state the reduction and assemble the bound, then construct the auxiliary market and implement its prices.

\subsection{The reduction and the revenue bound}\label{sec:assembly}

Fix integers $k,L\ge1$, $\eps\ge0$, a market $F\in\cF_k(\eps)$, and $0<s<1/2$. Abbreviate $\Rstar:=\Rstar(F)$ and $\LPP_L:=\LPP_L(F)$. The parameter $s$ controls the reliability of the competitive threshold constructed below; $\delta:=s/(1-s)$ bounds the resulting relative shortfall in the revenue benchmark.

\begin{proposition}[Reduction to the auxiliary market]\label{prop:reduction}
There is an auxiliary instance $I=I(F,s)$ with capacity $K\le k$ such that:
\begin{enumerate}[label=(\roman*),leftmargin=*]
\item Capacity suffices to collect all budgets at individual values, as in~\eqref{eq:auxiliary-affordable}, and the unrestricted auxiliary revenue $B:=\LPP^X_\infty(I)$ obeys
\[
\Rstar\le B+\delta\Rstar.
\]
\item Every feasible price plan for $I$ using at most $L$ prices, with revenue $b$ and inventory requirement $N$, admits a deterministic $L$-level mechanism for $F$ with expected revenue at least
\[
c_k\left(b-\sqrt{\frac{\eps\Rstar}{s}bN}\right).
\]
\end{enumerate}
\end{proposition}

Part~(i) compares unrestricted auxiliary revenue $B$ with optimal mechanism revenue $\Rstar$. The auxiliary guarantee in Proposition~\ref{thm:auxiliary} then bounds the cost of using only $L$ prices, and part~(ii) transfers those prices back to the original buyers. Writing $V_L:=\LPP^X_L(I)$ and using $N\le K\le k$ for an optimal plan gives
\begin{equation}\label{eq:comparison-chain}
\begin{aligned}
\Rstar&\le B+\delta\Rstar &&\text{(revenue benchmark)},\\
B&\le C_L^\star V_L &&\text{(coarse prices)},\\
\LPP_L&\ge c_k\left(V_L-\sqrt{\frac{k\eps\Rstar}{s}V_L}\right)
&&\text{(implementation)}.
\end{aligned}
\end{equation}
Only the middle comparison loses revenue because the price list is short. The first and third comparisons control the reduction errors: a smaller $s$ brings the benchmark closer to optimal revenue but weakens the implementation bound. Balancing these errors gives the finite-market guarantee.

\begin{proof}[Proof of Proposition~\ref{prop:finite-bound}]
Take any $0<s<1/2$ with $s/(1-s)+\sqrt{C_L^\star k\eps/s}<1$. The first two comparisons in~\eqref{eq:comparison-chain} give
\[
V_L\ge\frac{B}{C_L^\star}\ge\frac{1-\delta}{C_L^\star}\Rstar.
\]
The condition on $s$ makes the implementation lower bound positive and increasing in $V_L$ over this range. Substituting the displayed lower bound for $V_L$ into the third comparison gives
\[
\begin{aligned}
\LPP_L
&\ge c_k V_L\left(1-\sqrt{\frac{k\eps\Rstar}{sV_L}}\right)\\
&\ge\frac{c_k}{C_L^\star}\Rstar
\left(1-\delta-\sqrt{\frac{C_L^\star k\eps}{s}(1-\delta)}\right)\\
&\ge\frac{c_k}{C_L^\star}\Rstar
\left(1-\delta-\sqrt{\frac{C_L^\star k\eps}{s}}\right).
\end{aligned}
\]
Thus
\[
\frac{\Rstar(F)}{\LPP_L(F)}
\le\frac{C_L^\star}{c_k}
\frac1{1-s/(1-s)-\sqrt{C_L^\star k\eps/s}}.
\]
For $0<\eps\le\eps_0(k)$, choose $s=(k\eps)^{1/3}$. Appendix~\ref{app:finite-bound} verifies the condition on $s$ and bounds the last expression by $C_L^\star/c_k+8(2+\sqrt2)(k\eps)^{1/3}$. For $\eps=0$, take $s\downarrow0$ in the displayed bound.
\end{proof}

\subsection{Constructing the auxiliary benchmark}\label{sec:benchmark}\label{sec:pieces}\label{sec:prelim-revenue}\label{sec:localization}

We construct the auxiliary instance in Proposition~\ref{prop:reduction} and prove part~(i). A competitive threshold provides the inventory charge in the ex-ante allocation and bounds the revenue forgone when that allocation leaves capacity unused.

Let $U_i$ be independent and uniform on $(0,1)$, write $U=(U_1,\ldots,U_n)$, and represent values by $v_i=v_i(U_i)$. For regular buyers, marginal revenue is $\varphi_i(q):=R_i'(q)$, defined almost everywhere and nonincreasing. Write $Y_i:=\max\{\varphi_i(U_i),0\}$ for buyer $i$'s nonnegative marginal revenue, and $Y=(Y_1,\ldots,Y_n)$. The optimal revenue is the expected sum of the $k$ largest coordinates of $Y$, padding with zeros if necessary, by the virtual-surplus representation of \citet{Myerson81}. Appendix~\ref{app:threshold-allocation} gives the quantile conventions and the representation at atoms.

The marginal revenue at the inventory cutoff is $Z$, the $k$-th largest coordinate of $Y$, again padding with zeros. Choose the competitive threshold
\[
a:=\sup\{z\ge0:\Pr_U(Z\ge z)\ge1-s\}.
\]
Competing marginal revenues usually meet the threshold $a$: $\Pr_U(Z\ge a)\ge1-s$, while $\Pr_U(Z\le a)\ge s$ (Lemma~\ref{lem:window}). The same threshold controls implementation losses in Section~\ref{sec:realization}.

Choose ex-ante service probabilities $x=(x_1,\ldots,x_n)$ to maximize revenue net of inventory valued at $a$ per unit:
\begin{equation*}
\max_{x\in[0,1]^n,\;\sum_{i=1}^n x_i\le k}
\sum_{i=1}^n\bigl(R_i(x_i)-ax_i\bigr).
\end{equation*}
Among maximizers, choose one with the largest total quantity, and break remaining ties deterministically. Concavity and continuity ensure existence. The total expected allocation is
\[
K:=\sum_{i=1}^n x_i.
\]
Buyer $i$'s selected ex-ante revenue is $r_i:=R_i(x_i)$. For $x_i>0$, her reference price $w_i:=r_i/x_i=v_i(x_i)$ is this revenue per expected unit allocated. The allocation is chosen at the common inventory cost $a$, so the reference price need not be her monopoly price. Concavity and optimality imply $\varphi_i(q)\ge a$ for almost every $0<q<x_i$: the allocation stops before marginal revenue falls below the opportunity cost of inventory.

The allocation satisfies $w_i\ge a$ and $D_i(a)\ge x_i$ whenever $x_i>0$. Its unused capacity has small opportunity cost:
\[
a(k-K)\le\frac{s}{1-s}\Rstar.
\]
When $a>0$, with probability at least $1-s$ there are $k$ buyers whose marginal revenues meet the threshold, ensuring $K\ge k(1-s)$. When $a=0$, unused capacity has no opportunity cost. Lemma~\ref{lem:contact} proves these properties. Choosing the largest total quantity among maximizers also includes useful revenue opportunities with zero markup above $a$.

For every buyer with $r_i>0$, create an auxiliary buyer whose \emph{budget} $\beta_i=r_i=R_i(x_i)$ is her selected ex-ante revenue and whose \emph{per-unit value} is her reference price $w_i$. Give the auxiliary seller \emph{capacity} $K$, the total expected allocation, which may be fractional and is at most the original stock $k$. Denote the resulting market by $I=I(F,s)$.

The values and budgets are deterministic functions of the original distributions. They do not add budget constraints to the original buyers. Buyers with $r_i=0$ are omitted; if no positive budget remains, the auxiliary instance is empty and its revenues are zero.

At her reference price, each auxiliary buyer spends her budget and uses $\beta_i/w_i=x_i$ units, matching the original ex-ante service probability. Thus collecting all auxiliary budgets requires no more than capacity $K$:
\[
\sum_{i:r_i>0}\frac{\beta_i}{w_i}
=\sum_{i:r_i>0}x_i\le K.
\]
The unrestricted auxiliary revenue $B$ is the sum of these selected ex-ante revenues:
\begin{equation*}
B:=\sum_{i=1}^n r_i
=\LPP^{X}_{\infty}(I).
\end{equation*}
The revenue comparison in part~(i) now follows from the ex-ante relaxation.

\begin{proof}[Proof of Proposition~\ref{prop:reduction}(i)]
For any feasible incentive-compatible mechanism, define its interim allocation at quantile $q\in(0,1)$ by
\[
g_i(q):=\E_{v_{-i}\sim F_{-i}}[X_i(v_i(q),v_{-i})],
\qquad y_i:=\int_0^1 g_i(q)\,dq.
\]
Here $y_i$ is buyer $i$'s ex-ante service probability, and $\sum_{i=1}^n y_i\le k$. Internal mechanism randomness is already averaged in $X_i$ and $P_i$. The marginal-revenue representation and monotonicity of $\varphi_i$ give
\[
\E_{v\sim F}[P_i(v)]\le\int_0^1\varphi_i(q)g_i(q)\,dq
\le\int_0^{y_i}\varphi_i(q)\,dq=R_i(y_i).
\]
Adding the nonnegative slack $a(k-\sum_{i=1}^n y_i)$ and using optimality of $x$,
\[
\Rstar\le ak+\sum_{i=1}^n(r_i-ax_i)
=B+a(k-K)\le B+\frac{s}{1-s}\Rstar.\qedhere
\]
\end{proof}

We next transfer feasible price plans for this auxiliary market to the original buyers.

\subsection{Implementing the selected prices}\label{sec:realization}

To prove part~(ii) of Proposition~\ref{prop:reduction}, fix any feasible price plan using at most $L$ prices, with revenue $b$ and inventory requirement $N\le K$. Its prices are at least $a$, the lower bound on auxiliary values. The auxiliary inventory requirement $r_i/p$ need not bound the original buyer's acceptance probability $D_i(p)$, as Example~\ref{ex:demand-bill} illustrates. We first bound the expected revenue shortfall from the assigned prices, then account for uncertain purchases under the stock constraint.

Individual replaceability limits each buyer's revenue above the competitive threshold, which controls the expected payment lost when her price is lowered.

The opportunity cost of serving buyer $i$, measured in marginal revenue, is $T_i$, the $k$-th largest nonnegative virtual value among the other buyers, padding with zeros. Write $U_{-i}=(U_j)_{j\ne i}$. The pivotal identity expresses buyer $i$'s removal loss as $\Delta_i(F)=\E_U[(Y_i-T_i)_+]$: only her marginal revenue above the available substitute contributes to this loss (Lemma~\ref{lem:pivotal}). The following lemma bounds her revenue net of the inventory charge $aq$.

\begin{lemma}[Localization and the revenue ceiling]\label{lem:localization}
Every buyer satisfies
\begin{equation*}
\max_{q\in[0,1]}\{R_i(q)-aq\}
=\E_{U_i}[(Y_i-a)_+]\le\frac{\eps}{s}\Rstar.
\end{equation*}
\end{lemma}

\begin{proof}
Lemma~\ref{lem:window} gives $\Pr_{U_{-i}}(T_i\le a)\ge s$. Independence and the large-market assumption imply
\[
\eps\Rstar\ge\Delta_i(F)
\ge\Pr_{U_{-i}}(T_i\le a)\E_{U_i}[(Y_i-a)_+]
\ge s\E_{U_i}[(Y_i-a)_+].
\]
For the equality, $R_i(q)-aq=\int_0^q(\varphi_i(t)-a)\,dt$. The integrand is nonincreasing, so its cumulative integral is maximized by including its positive part. That part integrates to $\E_{U_i}[(Y_i-a)_+]$, since $a\ge0$.
\end{proof}

For a buyer with $r_i>0$, define her positive revenue shortfall by $d_i(p):=(r_i-pD_i(p))_+$.

\begin{lemma}[Individual revenue shortfall]\label{lem:shortfall}
For every buyer with $r_i>0$ and every $0<p\le w_i$ with $p\ge a$,
\begin{equation*}
d_i(p)\le\min\left\{\frac{\eps}{s}\Rstar,\frac{r_i^2}{p}\right\}.
\end{equation*}
\end{lemma}

For the first bound, $p\le w_i$ implies $D_i(p)\ge x_i$, so a price no lower than $a$ preserves expected revenue of at least $ax_i$. Only the excess $r_i-ax_i$, bounded by Lemma~\ref{lem:localization}, can be lost. The second bound follows from concavity of the buyer's revenue curve and bounds the loss by her selected ex-ante revenue times the auxiliary inventory requirement $r_i/p$. Appendix~\ref{app:shortfalls} proves both bounds.

To aggregate them, index the plan's nonempty blocks by $\ell$ and let $\lambda_{i\ell}$ be the fraction of buyer $i$'s budget in block $\ell$, with price $p_\ell>0$. These budget shares will become the probabilities with which she receives each price. The sums below range over buyers with $r_i>0$ and these nonempty blocks; set $\lambda_{i\ell}=0$ when no budget portion is assigned. For each buyer, $\sum_\ell\lambda_{i\ell}\le1$; whenever $\lambda_{i\ell}>0$, $a\le p_\ell\le w_i$. Summing the selected spending and its inventory requirements gives
\[
b=\sum_{i,\ell}\lambda_{i\ell}r_i,
\qquad N=\sum_{i,\ell}\lambda_{i\ell}\frac{r_i}{p_\ell}.
\]
Using $\min\{A,B\}\le\sqrt{AB}$ and Cauchy--Schwarz gives the aggregate shortfall
\begin{equation}\label{eq:aggregate-shortfall}
\mathcal E:=\sum_{i,\ell}\lambda_{i\ell}d_i(p_\ell)
\le\sqrt{\frac{\eps\Rstar}{s}}\sum_{i,\ell}\lambda_{i\ell}\sqrt{r_i\frac{r_i}{p_\ell}}
\le\sqrt{\frac{\eps\Rstar}{s}bN}.
\end{equation}
Both bounds matter. Revenue opportunities using little auxiliary inventory have limited relative loss. Those using more inventory have small absolute loss, and their combined inventory requirement is bounded by $K$.

Assign buyer $i$ price $p_\ell$ with probability $\lambda_{i\ell}$, and no offer with the remaining probability, independently across buyers and before values are observed. Each buyer receives at most one offer. Actual demand may exceed $N$. For comparison, form hypothetical payments by independently discarding some acceptances before applying the stock constraint. Conditional on assigning buyer $i$ price $p$, the positive-payment probability is $\min\{D_i(p),r_i/p\}$ and the expected payment is $\min\{pD_i(p),r_i\}$. The comparison therefore has total expected demand at most $N\le k$ and total expected payment $b-\mathcal E$. Appendix~\ref{app:assignments} gives the construction.

Actual offers serve every accepting buyer while inventory remains, in decreasing price order. Their potential payments dominate the independent comparison payments, so the correlation-gap bound in Lemma~\ref{lem:cg} guarantees expected revenue at least $c_k(b-\mathcal E)$. Substituting~\eqref{eq:aggregate-shortfall} gives the bound in part~(ii) of Proposition~\ref{prop:reduction}. Every lottery outcome uses at most the plan's $L$ prices, so some fixed assignment earns at least the lottery average. Fixing it before values are drawn gives the required deterministic $L$-level mechanism.

\section{Markets Approaching the Sharp Bound}\label{sec:coarseness-constants}

We prove Proposition~\ref{prop:market-sharpness} by constructing private-value markets that reproduce the auxiliary inventory trade-off. Fix an auxiliary capacity $K>0$, a baseline value $a>0$, and finitely many premium types $t=1,\ldots,T$, with budgets $r_t>0$ and per-unit values $w_t>a$. Collecting their budgets at individual values requires $Q:=\sum_t r_t/w_t\le K$ units. Add a baseline budget $a(K-Q)$ at value $a$, which uses the remaining capacity when charged that value. The resulting auxiliary profile $u$ has unrestricted revenue
\[
B:=\sum_{t=1}^T r_t+a(K-Q)
\]
and $L$-price revenue $V_L(u,K)$. Lemma~\ref{lem:finite-covering-profiles} shows that such profiles can approach $C_L^\star$ arbitrarily closely.

For stock $k$, create $m$ independent buyers of each premium type, with truncated equal-revenue demand $D_{t,m}(p)=c_t/p$ on $[c_t,w_t]$. Here $c_t:=kr_t/(Km)$ is each buyer's expected payment before the stock constraint at every price in that interval. Choose $m$ large enough that $c_t<a$ for every type, and add $k+1$ deterministic buyers of value $a$. At any offer $a\le p\le w_t$, a type's total expected spending and demand before the stock constraint are
\[
mpD_{t,m}(p)=\frac{k}{K}r_t,
\qquad mD_{t,m}(p)=\frac{k}{K}\frac{r_t}{p}.
\]
Lowering an eligible price preserves expected spending and increases expected inventory use, exactly as in the auxiliary market. The baseline buyers provide a sale opportunity for stock left after the premium offers.

Spreading each type's spending across more buyers makes every premium buyer replaceable. Removing a baseline buyer changes nothing because $k$ others remain. At each fixed stock, the resulting market $F_{k,m}$ is $\eps_m$-large with $\eps_m=\max_t r_t/(Kam)\to0$. This expansion leaves aggregate demand uncertain. Increasing stock afterward makes the revenue effect of those fluctuations small relative to total revenue.

Personalized prices $w_t$, followed by baseline sales, nearly collect unrestricted auxiliary revenue. Conversely, any assignment of $L$ prices can be converted into a feasible auxiliary plan whose revenue is at least the assignment's normalized expected revenue. Writing $w_{\max}:=\max_t w_t$, Proposition~\ref{prop:sharpness-transfer} in Appendix~\ref{app:sharpness-transfer} establishes
\begin{equation*}
\LPP_\infty(F_{k,m})\ge\frac{k}{K}B-w_{\max}\sqrt{k},
\qquad
\LPP_L(F_{k,m})\le\frac{k}{K}V_L(u,K).
\end{equation*}
The upper bound covers every assignment of the selected prices, including assignments whose realized demand exceeds stock. The lower bound is achieved within personalized posted pricing, so the resulting loss measures the restriction on price differentiation under either benchmark.

\begin{proof}[Proof of Proposition~\ref{prop:market-sharpness}]
Fix $L\ge1$ and $\eta>0$. Lemma~\ref{lem:finite-covering-profiles} supplies a finite positive profile of the preceding form with
\[
\frac{B}{V_L(u,K)}>C_L^\star-\frac\eta2.
\]
Keep this profile fixed and choose an integer $k_0\ge1$ such that
\[
\frac{K w_{\max}}{V_L(u,K)\sqrt{k_0}}\le\frac\eta2.
\]
For every $k\ge k_0$ and $\eps>0$, choose $m$ large enough that $c_t<a$ for all premium types and $\eps_m\le\eps$. Proposition~\ref{prop:sharpness-transfer} gives a market $F_{k,m}\in\cF_k(\eps)$ satisfying
\[
\frac{\LPP_\infty(F_{k,m})}{\LPP_L(F_{k,m})}
\ge\frac{B}{V_L(u,K)}-\frac{K w_{\max}}{V_L(u,K)\sqrt{k}}
>C_L^\star-\eta.\qedhere
\]
\end{proof}

\section{Discussion}\label{sec:discussion}

The revenue value of additional prices comes from reducing the inventory cost of pooling buyers. In the auxiliary market, a short price list balances this cost against the spending forgone by excluding some demand. The same trade-off determines the sharp revenue ratios in the original market as buyers become individually replaceable at each fixed stock and stock then grows. The two revenue benchmarks give the same limiting cost, isolating the restriction on price differentiation.

A cap on the number of prices leaves scope for targeting through buyer assignments and control over service priority. Assignments may use detailed customer information even when the price list is short. The preferred number of prices still depends on demand and implementation costs. Revenue guarantees alone do not determine the effects on consumer surplus or access.

Targeted reservation offers illustrate how these choices can be implemented. For general-admission tickets to an event with fixed capacity, a seller could use verified customer information to assign a few rates and collect purchase requests during a common window. Requests become binding when the window closes; scarce units go to accepting buyers with the highest assigned prices, and only those served pay. This gives the seller the same revenue as sequential offers in decreasing price order. Buyers may arrive at different times, provided the seller can defer allocation. Ticket-request systems illustrate the feasibility of collecting commitments before allocating stock.\footnote{Ticketmaster collects requests during a common window, followed by fulfillment and charging of successful requests. Requests may be canceled during the window, but not afterward. See \href{https://help.ticketmaster.com/hc/en-us/articles/40436800695953-How-do-I-request-tickets-for-my-event}{How do I request tickets for my event?} (accessed October 31, 2026). This documents deferred allocation, not allocation by assigned price.}

Alternatively, the seller could open offers to higher-price groups first and release remaining capacity to lower-price groups. Both implementations require enforceable assignments: buyers cannot switch to another customer's price or obtain a later discount by rejecting their offer. Coarse pricing alone need not entail this service priority: Sephora's deepest-discount tier receives earlier access. Whether comparable guarantees hold for immediate sales in an externally determined arrival order remains open.

The finite-market bound also leaves open how much competition is needed relative to inventory. Replicated customer groups growing faster than stock provide a sufficient condition, while the harmonic-value example in Remark~\ref{rem:nested} shows why small individual removal losses alone do not suffice. Neither the cube-root rate nor its coefficient is claimed optimal. The exact constant $C_L^\star$ remains unknown for $L\ge3$; the current bounds differ by $1/[(2L-2)(2L-1)]$. At fixed stock, establishing whether $C_L^\star/c_k$ is sharp would require a construction capturing both the inventory cost of coarse prices and uncertain sales.

\appendix
\numberwithin{lemma}{section}
\numberwithin{corollary}{section}
\numberwithin{definition}{section}
\numberwithin{remark}{section}
\numberwithin{proposition}{section}

\section{Auxiliary Pricing and Sharp Constants}\label{app:covering}

This appendix completes the proof of Proposition~\ref{thm:auxiliary}. We first justify the price plans introduced in Section~\ref{sec:apply-auxiliary}, then characterize the sharp ratio through inventory savings that a short price list fails to capture. Evaluating this loss gives the constants and lower-bound examples. The same construction supplies the finite positive profiles used in Section~\ref{sec:coarseness-constants}.

\subsection{Unrestricted pricing and canonical plans}

For fixed assigned prices, let $A=\{i:0<p_i\le w_i\}$ be the buyers with acceptable offers. Decreasing-price service attains
\begin{equation*}
\max\left\{\sum_{i\in A}y_i:
0\le y_i\le\beta_i,\quad
\sum_{i\in A}\frac{y_i}{p_i}\le K\right\},
\end{equation*}
where $y_i$ is the payment collected from buyer $i$. This program collects spending at the highest prices first, fully spending each budget until at most one final budget is partially collected. In particular, at fixed prices revenue is nondecreasing in every budget.

\begin{lemma}[Unrestricted auxiliary pricing]\label{lem:auxiliary-full}
The unrestricted auxiliary revenue is
\begin{equation*}
\LPP^{X}_{\infty}(I)
=\max\left\{\sum_{i=1}^n w_iq_i:
\sum_{i=1}^n q_i\le K,\quad
0\le q_i\le\frac{\beta_i}{w_i}\right\}.
\end{equation*}
An optimal policy charges each buyer her value and serves buyers in decreasing value order, fully collecting budgets until stock runs out, with at most one partially collected budget. If~\eqref{eq:auxiliary-affordable} holds, this revenue is $B=\sum_{i=1}^n\beta_i$.
\end{lemma}

\begin{proof}
If a policy collects payment $y_i$ from buyer $i$ at an acceptable price, it uses at least $y_i/w_i$ units and collects at most $\beta_i$. Hence $q_i=y_i/w_i$ is feasible for the displayed program, whose objective equals the policy's revenue. Conversely, the program allocates each unit to the highest remaining value until the corresponding bound $\beta_i/w_i$ is attained. Charging buyers their values in this order implements that allocation. Under~\eqref{eq:auxiliary-affordable}, every upper bound can be attained.
\end{proof}

\begin{proof}[Proof of Lemma~\ref{lem:auxiliary-plans}]
Take a feasible plan. For every buyer with a selected budget portion, assign the highest block price intersecting that portion; assign no offer to other buyers. Each assigned price is at most the buyer's value. Moving all of her selected spending to this price weakly reduces the inventory requirement. These selected payments therefore give a feasible allocation of at most $N\le K$ units, with total revenue $b$, at the assigned prices. Actual desired quantities use full budgets and weakly exceed these selected quantities. Serving buyers in decreasing assigned-price order maximizes revenue subject to their desired quantities and stock $K$. It therefore earns at least $b$. In particular, splitting a budget in the plan does not require multiple offers to that buyer.

Conversely, consider any auxiliary policy earning $b$ and using at most $K$ units. For every buyer from whom it collects a positive payment, move that payment to the highest price in the policy's price list that does not exceed her value. This weakly lowers its inventory requirement. At these highest eligible prices, inventory per budget dollar is nondecreasing in the value order. Collecting the first $b$ dollars in that order therefore requires no more inventory than the original payments. The highest eligible price is constant on each of at most $L$ consecutive blocks of this prefix. Raise each nonempty block's price to the value at its last budget dollar; every buyer in the block still accepts it, and its inventory requirement weakly falls. This is a feasible plan with revenue $b$, so $b\le V_L(u,K)$. The two inequalities prove the claim.
\end{proof}

\subsection{General profiles and the sharp ratio}\label{sec:grouping-costs}

We extend Definition~\ref{def:plan} to finite-valued, nonnegative, nondecreasing, left-continuous profiles $u$ on $(0,B]$, setting $u(0)=u(0^+)$. The same boundary, revenue, and inventory requirement formulas define a plan. Repeated boundaries give empty blocks, which contribute zero and need no price evaluation. A zero unit cost is allowed as a limiting device. For $B=0$, put $u(0)=0$ and $V_L=0$. A profile with $B,K>0$ is \emph{admissible} if $\int_0^B u(R)\,dR\le K$.

\begin{remark}[Attainment]\label{rem:attainment}
The weak partitions $0\le R_1\le\cdots\le R_L\le B$ form a compact simplex. Since $u$ is nonnegative, nondecreasing and left-continuous, their inventory requirement $N=\sum_{\ell=1}^L u(R_\ell)(R_\ell-R_{\ell-1})$, with $R_0=0$, is lower semicontinuous. Thus the feasible set $\{N\le K\}$ is closed and the largest endpoint $V_L$ is attained. Empty blocks cause no difficulty because their contribution is zero.
\end{remark}

To measure the inventory savings captured by a short price list, let $\theta:(0,1]\to[0,1]$ be nonincreasing and left-continuous, and extend it to zero by its right limit. Interpret $1-\theta(y)$ as inventory per dollar: $\theta(y)$ is the saving relative to a common unit cost of one. Total savings at individual values are $A$. With $j\ge1$ price blocks followed by a final block at unit cost one, the largest savings are $S_j(\theta)$:
\begin{equation*}
A:=\int_0^1\theta(y)\,dy,
\qquad
S_j(\theta):=\max_{0=y_0\le\cdots\le y_j\le1}
\sum_{\ell=1}^j\theta(y_\ell)(y_\ell-y_{\ell-1}).
\end{equation*}
For $j=0$, set $S_0(\theta):=0$. The maximum is attained because its objective is upper semicontinuous on the compact simplex. Savings not captured by these blocks or in the omitted suffix contribute to $A-S_j(\theta)$. The largest such loss is
\[
E_j:=\sup_\theta\{A-S_j(\theta)\},\qquad j=0,1,2,\ldots,
\]
where the supremum is over the profiles just specified and $A$ is the area of that same profile. In particular, $E_0=1$.

\begin{proposition}[Characterization of the sharp auxiliary ratio]\label{prop:det-tight}
The sharp auxiliary constant satisfies
\begin{equation}\label{eq:det-characterization}
C_L^\star=1+E_{L-1}
=1+\sup_\theta\{A-S_{L-1}(\theta)\}.
\end{equation}
For $L\ge3$,
\begin{equation*}
1+\frac1{2L-1}\le C_L^\star\le1+\frac1{2(L-1)}.
\end{equation*}
The lower endpoint need not be exact. In particular,
\begin{equation*}
C_3^\star\ge1+\frac{277}{1350}>1+\frac15.
\end{equation*}
\end{proposition}

\begin{proof}[Proof of \eqref{eq:det-characterization}]
For any finite auxiliary instance, restrict each budget to the payment collected under unrestricted auxiliary pricing. This preserves unrestricted revenue and can only lower $L$-price revenue, while making all remaining budgets affordable at individual values. By Lemmas~\ref{lem:auxiliary-full} and~\ref{lem:auxiliary-plans}, it therefore suffices for the upper bound to compare $B$ with $V_L(u,K)$ for an admissible profile.

The following comparison applies to every such profile. Set $j:=L-1$ and $b:=B/(1+E_j)$. If $u(b)=0$, one block earns revenue $b$ with zero inventory requirement. Otherwise normalize the unit costs on this prefix by
\[
\theta(z):=1-\frac{u(bz)}{u(b)},\qquad 0<z\le1.
\]
Choose a prefix attaining $S_j(\theta)$, multiply its endpoints by $b$, and append endpoint $b$. This uses at most $L$ blocks, with inventory requirement
\[
\begin{aligned}
N&=b\,u(b)\bigl(1-S_j(\theta)\bigr)\\
&=\int_0^b u(R)\,dR+b\,u(b)\bigl(A-S_j(\theta)\bigr)\\
&\le\int_0^b u(R)\,dR+(B-b)u(b)
\le\int_0^B u(R)\,dR\le K.
\end{aligned}
\]
Thus $V_L(u,K)\ge B/(1+E_{L-1})$, proving $C_L^\star\le1+E_{L-1}$.

For the reverse inequality, fix $\theta$ and write $S:=S_{L-1}(\theta)$. A zero loss $A-S$ contributes only the trivial ratio one. If $A-S>0$, then $S<1$. Give the first dollar of spending individual unit costs $1-\theta(R)$, then append $A-S$ dollars at unit cost one. Set capacity equal to the inventory needed at individual values:
\begin{equation}\label{eq:covering-extremal-embedding}
B:=1+A-S,\qquad K:=1-S,\qquad
u(R):=\begin{cases}
1-\theta(R),&0<R\le1,\\
1,&1<R\le B.
\end{cases}
\end{equation}
A prefix attaining $S$, followed by endpoint one, uses at most $L$ blocks and requires at most $1-S=K$ units, so $V_L\ge1$. Conversely, a plan ending at $b>1$ has a final block of unit cost one. Merge any other blocks ending above one into that block. The preceding endpoints number at most $L-1$ and lie in $[0,1]$. If their savings under $\theta$ are $\sigma$, the inventory requirement is $b-\sigma\ge b-S>K$. Thus $V_L=1$ and $B/V_L=1+A-S$.

To realize this ratio with finite auxiliary buyers, choose $\delta\in(0,1/2)$ and an integer $m\ge1$, and put
\[
\widetilde\theta(y):=\delta+(1-2\delta)\frac{\lfloor m\theta(y)\rfloor}{m},
\qquad
\widetilde A:=\int_0^1\widetilde\theta(y)\,dy.
\]
This is a nonincreasing, left-continuous finite-step profile with values in $[\delta,1-\delta]$ and uniform distance from $\theta$ at most $\delta+1/m$. Every prefix has length at most one, so both its integral and $S_{L-1}$ change by at most $\delta+1/m$. Hence
\[
\left|\bigl(\widetilde A-S_{L-1}(\widetilde\theta)\bigr)-(A-S)\right|
\le2(\delta+1/m).
\]
Apply~\eqref{eq:covering-extremal-embedding} to $\widetilde\theta$. For a sufficiently fine approximation, its final interval has positive length. Its initial unit costs lie strictly between zero and one, and the same argument gives $V_L=1$. Each constant interval defines a buyer with budget equal to its length and value equal to the reciprocal of its unit cost. The resulting finite auxiliary revenue ratios approach $1+A-S$. Taking the supremum proves~\eqref{eq:det-characterization}. The bounds and the three-price improvement are established below.
\end{proof}

\begin{proposition}[Profile representation of the sharp ratio]\label{prop:auxiliary-sharp}
The auxiliary constant satisfies
\begin{equation*}
C_L^\star=\sup_{\substack{(B,K,u)\\\mathrm{admissible}}}
\frac{B}{V_L(u,K)}.
\end{equation*}
\end{proposition}

\begin{proof}
The upper comparison in the preceding proof bounds every admissible profile's ratio by $1+E_{L-1}=C_L^\star$. The finite positive profiles constructed there are admissible and approach that bound. This proves the representation. For every positive-loss profile of the form~\eqref{eq:covering-extremal-embedding}, the same construction gives finite positive profiles with $V_L=1$ and revenue ratios converging to its ratio.
\end{proof}

\subsection{Evaluating the constants}

\begin{lemma}[Prefix approximation]\label{lem:prefix-grid}
For every $j\ge1$,
\begin{equation*}
A-S_j(\theta)\le\frac1{2j}.
\end{equation*}
For one block, the exact worst-case loss is
\begin{equation*}
E_1=\frac1e.
\end{equation*}
\end{lemma}

\begin{proof}
Represent the decreases of $\theta$ by the finite positive Borel measure $\mu$ on $[0,1]$ satisfying $\theta(y)=\mu([y,1])$ for $y\in[0,1]$. For endpoints $0=y_0\le y_1\le\cdots\le y_j\le1$, write $P=\{y_1,\ldots,y_j\}$. For $\tau\in[0,1]$, let $\operatorname{pre}_P(\tau)$ be the last endpoint weakly below $\tau$, or zero if there is none. Tonelli's theorem gives
\begin{equation*}
A-\sum_{\ell=1}^j\theta(y_\ell)(y_\ell-y_{\ell-1})
=\int_{[0,1]}\bigl(\tau-\operatorname{pre}_P(\tau)\bigr)\,\mu(d\tau).
\end{equation*}
Thus each downward drop loses its size times the distance back to the preceding endpoint, including losses in the unselected suffix.

Put $\Delta:=1/j$ and use the endpoints $\xi,\xi+\Delta,\ldots,\xi+(j-1)\Delta$, with $\xi$ uniform on $[0,\Delta]$. The profile and measure are fixed. For $\tau\in[0,\Delta)$, the expected backward distance is $\tau-\tau^2/(2\Delta)\le\Delta/2$. For $\tau\in[\Delta,1]$, it is $\Delta/2$. Since $\mu([0,1])\le1$, the expected loss is at most $1/(2j)$. Some fixed shift therefore achieves this bound.

For the sharper one-block value, put $r:=S_1(\theta)=\max_{0\le y\le1}y\theta(y)$. If $r=0$, the loss is zero. Otherwise $\theta(y)\le\min\{1,r/y\}$, so
\[
A\le r+\int_r^1\frac r y\,dy
=r+r\log\frac1r.
\]
Hence $A-S_1(\theta)\le r\log(1/r)\le1/e$. Equality holds for
\begin{equation}\label{eq:equal-revenue-profile}
\theta(y)=\min\left\{1,\frac{1}{ey}\right\},\qquad 0<y\le1.\qedhere
\end{equation}
\end{proof}

\paragraph{One price.}
Since $E_0=1$, equation~\eqref{eq:det-characterization} gives $C_1^\star=2$. Concretely, take $u=0$ on $(0,1]$ and $u=1$ on $(1,2]$, with $K=1$. Unrestricted revenue is two, while any single-block plan earns at most one.

\paragraph{Two prices.}
Since $E_1=1/e$, equation~\eqref{eq:det-characterization} gives $C_2^\star=1+1/e$. The profile~\eqref{eq:equal-revenue-profile} has $A=2/e$ and $S_1=1/e$. Its embedding~\eqref{eq:covering-extremal-embedding} has $B=1+1/e$, $K=1-1/e$, and $V_2=1$. The finite positive approximation above makes both examples limits of finite auxiliary markets.

\subsection{Lower bounds for the covering constant}

\paragraph{The linear-profile illustration.}\label{app:linear-profile}
For $L\ge3$, consider the profile used in Section~\ref{sec:apply-auxiliary}. Take $\theta(y)=1-y$ and put $j:=L-1$. A $j$-block prefix with endpoint $b$ and widths $h_\ell:=y_\ell-y_{\ell-1}$ has savings
\[
\begin{aligned}
\sum_{\ell=1}^j(1-y_\ell)h_\ell
&=b-\frac12\left(b^2+\sum_{\ell=1}^j h_\ell^2\right)\\
&\le b-\frac{j+1}{2j}b^2
\le\frac{j}{2(j+1)}.
\end{aligned}
\]
Equal widths $h_\ell=1/(j+1)$ attain both inequalities. Thus $A=1/2$ and $S_{L-1}=(L-1)/(2L)$. The construction~\eqref{eq:covering-extremal-embedding} gives
\[
u(R)=\min\{R,1\},\qquad
B=1+\frac1{2L},\qquad K=\frac12+\frac1{2L},\qquad V_L=1.
\]
Proposition~\ref{prop:auxiliary-sharp} justifies the finite positive approximations used in the main text.

\paragraph{The general lower bound.}
Put $j:=L-1\ge1$, $\alpha:=1/(2j+1)$, and
\[
\theta(y):=1-(y-\alpha)_+.
\]
For a $j$-block prefix $0=y_0\le y_1\le\cdots\le y_j=b\le1$, set $h_\ell:=y_\ell-y_{\ell-1}$ for $\ell=1,\ldots,j$. Its lower step area $\sigma:=\sum_{\ell=1}^j\theta(y_\ell)h_\ell$ satisfies
\begin{align*}
\sigma
&\le(1+\alpha)b-\sum_{\ell=1}^j y_\ell h_\ell\\
&=(1+\alpha)b-\frac12\left(b^2+\sum_{\ell=1}^j h_\ell^2\right)\\
&\le(1+\alpha)b-\frac{j+1}{2j}b^2.
\end{align*}
The identity follows by telescoping $y_\ell^2-y_{\ell-1}^2$, and the last inequality is Cauchy--Schwarz. This quadratic is maximized at $b=2j\alpha=1-\alpha$. Equal blocks of width $2\alpha$ attain every inequality, giving
\[
S_j(\theta)=\frac{j}{2(j+1)}(1+\alpha)^2,
\qquad A=1-\frac{(1-\alpha)^2}{2},
\qquad A-S_j(\theta)=\frac1{2L-1}.
\]
Together with Lemma~\ref{lem:prefix-grid}, this proves the bounds for $L\ge3$ in Proposition~\ref{prop:det-tight}.

\paragraph{A three-price improvement.}
For $L=3$, take
\begin{equation*}
\theta(y)=\begin{cases}
1,&0\le y\le\tfrac15,\\[2pt]
\dfrac{11-10y}{9},&\tfrac15\le y\le\tfrac45,\\[6pt]
\dfrac{13-10y}{15},&\tfrac45\le y\le1.
\end{cases}
\end{equation*}
Its integral is $A=49/75$. For two endpoints $0\le s\le t\le1$, the savings are
\[
F(s,t):=\theta(s)s+\theta(t)(t-s)
=t\theta(t)+s\bigl(\theta(s)-\theta(t)\bigr).
\]
If $t\le1/5$, then $F(s,t)=t\le1/5$. For fixed $t$, the objective decreases in $s$ above $4/5$, because its derivative there is $(13-20s)/15-\theta(t)<0$. We may therefore restrict $s\le4/5$. On this interval, $\theta(s)\le(11-10s)/9$. Completing the square in $s$ gives, for $t\ge1/5$,
\[
\begin{aligned}
F(s,t)&\le t\theta(t)+\frac{(11-9\theta(t))^2}{360}\\
&=\begin{cases}
\dfrac{22t-15t^2}{18},&\tfrac15\le t\le\tfrac45,\\[6pt]
\dfrac{256+8760t-5100t^2}{9000},&\tfrac45\le t\le1.
\end{cases}
\end{aligned}
\]
These two quadratics have maxima $121/270$ and $683/1530$, respectively, with the first larger. The bound $121/270$ is attained at $(s,t)=(11/30,11/15)$. Hence $S_2=121/270$ and
\[
A-S_2(\theta)=\frac{49}{75}-\frac{121}{270}
=\frac{277}{1350}>\frac15,
\]
proving the three-price improvement in Proposition~\ref{prop:det-tight}.

\subsection{Finite profiles for the market lower bound}

\begin{lemma}[Finite positive profiles suffice]\label{lem:finite-covering-profiles}
For every $L\ge1$ and $\eta>0$, there is a finite-step profile with $u(R)\in(0,1)$ on $(0,1]$, followed by a nonempty interval of unit cost one, such that
\[
\int_0^B u(R)\,dR=K>0,
\qquad V_L(u,K)=1,
\qquad \frac{B}{V_L(u,K)}>C_L^\star-\eta.
\]
Thus restricting the supremum defining $C_L^\star$ to these profiles does not change it.
\end{lemma}

\begin{proof}
The calculations above give $E_{L-1}>0$ for every finite $L$. Choose $\theta$ with positive loss $A-S_{L-1}(\theta)$ sufficiently close to $E_{L-1}$. Let $\widetilde\theta$ be a finite-step approximation from the proof of Proposition~\ref{prop:det-tight}. Applying~\eqref{eq:covering-extremal-embedding} gives initial unit costs strictly between zero and one, a nonempty final interval of unit cost one, total inventory requirement $K=1-S_{L-1}(\widetilde\theta)>0$, and $V_L=1$. Choosing the approximation sufficiently fine makes the revenue ratio exceed $C_L^\star-\eta$.
\end{proof}

\section{Implementing Auxiliary Prices}\label{app:toolkit}

This appendix supplies the omitted verification for Section~\ref{sec:argument}. We first establish the threshold and allocation properties used to construct the benchmark, then verify the removal-loss and individual-shortfall arguments used in implementation. The payment coupling and correlation-gap bound complete the reduction. We finish with the finite-market constants and an example distinguishing auxiliary inventory requirement from actual demand.

\subsection{Marginal revenue and the ex-ante allocation}\label{app:threshold-allocation}

Let $U_i$ be independent and uniform on $(0,1)$ for $i=1,\ldots,n$, and write $U=(U_1,\ldots,U_n)$. Represent buyer $i$'s value by $v_i=v_i(U_i)$, which has distribution $F_i$ even at atoms. Expectations involving virtual values are taken over these quantiles. The finite-mean assumption gives $R_i(0^+)=0$.\footnote{For an event $A$, $\ind{A}$ equals one if $A$ occurs and zero otherwise. If $\operatorname{ess\,sup}v_i=\infty$ then $v_i(q)\uparrow\infty$ as $q\downarrow0$ and $R_i(q)\le\E_{v_i\sim F_i}[v_i\ind{v_i\ge v_i(q)}]\to0$ by dominated convergence; if $\operatorname{ess\,sup}v_i=M<\infty$ then $R_i(q)\le qM\to0$.}

For a distribution $F_i$ with positive density $f_i$ on an interval support, regularity is equivalent to monotonicity of the virtual value $v-\tfrac{1-F_i(v)}{f_i(v)}$. For regular $F_i$ the quantile-space \emph{marginal revenue}, or virtual value, is $\varphi_i(q):=R_i'(q)$, well defined for almost every $q$ and nonincreasing; we fix the right-derivative version on $[0,1)$ and set $\varphi_i(1):=\varphi_i(1^-)$, so that $R_i(q)=\int_0^q\varphi_i(u)\,du$. Write
\[
Y_i:=\max\{\varphi_i(U_i),0\},
\qquad Y=(Y_1,\ldots,Y_n).
\]
The positive part reflects the seller's ability to exclude buyers with negative marginal revenue.

For a nonnegative vector $y$, let $S_k(y)$ be the sum of its $k$ largest entries, padding with zeros if the vector has fewer than $k$ coordinates. The standard virtual-surplus representation for independent regular buyers gives
\[
\Rstar(F)=\E_U[S_k(Y)],
\]
and the same identity holds after removing any buyer \citep{Myerson81,HartlineBook}.

Let $Z$ be the $k$-th largest coordinate of $Y$, again padding with zeros. For $0<s<1/2$, use the competitive threshold from Section~\ref{sec:benchmark}:
\[
a:=\sup\{z\ge0:\Pr_U(Z\ge z)\ge1-s\}.
\]

\begin{lemma}[Threshold probabilities]\label{lem:window}
The threshold satisfies
\[
\Pr_U(Z\ge a)\ge1-s,
\qquad \Pr_U(Z\le a)\ge s,
\qquad ak(1-s)\le\Rstar.
\]
\end{lemma}

\begin{proof}
The defining set is nonempty because $Z\ge0$. It is bounded because $\E_U[S_k(Y)]=\Rstar$ and $S_k(Y)\ge kz$ on $\{Z\ge z\}$. Taking thresholds increasing to $a$ gives $\Pr_U(Z\ge a)\ge1-s$. Every $a+t$ with $t>0$ lies above the supremum, so $\Pr_U(Z<a+t)>s$. Letting $t\downarrow0$ gives $\Pr_U(Z\le a)\ge s$. Finally, $S_k(Y)\ge ka$ on $\{Z\ge a\}$ gives the last inequality. These weak inequalities also cover atoms at $a$.
\end{proof}

Let $x$, $K$, and $w_i$ be the ex-ante allocation, its total quantity, and the reference prices chosen in Section~\ref{sec:benchmark}.

\begin{lemma}[Ex-ante allocation and unused capacity]\label{lem:contact}
Every buyer with $x_i>0$ satisfies $w_i\ge a$ and $D_i(a)\ge x_i$. Moreover,
\begin{equation*}
K=\min\left\{k,\sum_{i=1}^n\Pr_{U_i}(\varphi_i(U_i)\ge a)\right\},
\qquad
 a(k-K)\le\frac{s}{1-s}\Rstar.
\end{equation*}
If $a>0$, then $K\ge k(1-s)$.
\end{lemma}

\begin{proof}
If $R_i(x_i)-ax_i<0$, setting $x_i=0$ would improve the objective. Hence $w_i\ge a$, and the weak-tail quantile convention gives $D_i(a)\ge D_i(w_i)\ge x_i$.

Let $t_i$ be the rightmost maximizer of $R_i(q)-aq$ on $[0,1]$. Concavity gives
\[
t_i=\Pr_{U_i}(\varphi_i(U_i)\ge a).
\]
No optimizer can have $x_i>t_i$, since reducing that coordinate to $t_i$ would improve its objective. If $K<k$ and some $x_i<t_i$, a small increase in $x_i$ either improves the objective or preserves it while increasing total quantity. Both contradict the choice of $x$. Thus either $K=k$ or $x_i=t_i$ for all $i$, proving the formula for $K$.

When $a>0$, the event $\{Z\ge a\}$ requires at least $k$ genuine buyers with $\varphi_i(U_i)\ge a$. Their expected number is at least $k(1-s)$, so the formula gives $K\ge k(1-s)$. Together with $ak(1-s)\le\Rstar$, this yields $a(k-K)\le aks\le s\Rstar/(1-s)$. When $a=0$, the same bound is immediate. Zero padding gives no lower bound on $K$ in that case.
\end{proof}

\subsection{Removal losses and individual revenue shortfalls}\label{app:shortfalls}

For each $i$, let $T_i$ be the $k$-th largest nonnegative virtual value among the other buyers, padding with zeros. Write $U_{-i}=(U_j)_{j\ne i}$. Thus $T_i$ is independent of $Y_i$, and $T_i\le Z$ implies
\[
\Pr_{U_{-i}}(T_i\le a)\ge\Pr_U(Z\le a)\ge s.
\]
This is the probability used in the proof of Lemma~\ref{lem:localization}.

\begin{lemma}[Pivotal identity]\label{lem:pivotal}
For every buyer $i$, $\Delta_i(F)=\E_U[(Y_i-T_i)_+]$.\footnote{For a real number $z$, $z_+:=\max\{z,0\}$.}
\end{lemma}

\begin{proof}
Inserting a coordinate $y_i$ into the other nonnegative virtual values increases their top-$k$ sum by $(y_i-T_i)_+$. Take expectations and use the virtual-surplus identity for the market with and without buyer $i$.
\end{proof}

\begin{proof}[Proof of Lemma~\ref{lem:shortfall}]
Since $p\le w_i$, we have $D_i(p)\ge x_i$. Thus $pD_i(p)\ge ax_i$, and Lemma~\ref{lem:localization} gives $d_i(p)\le r_i-ax_i\le\eps\Rstar/s$.

For the second bound, concavity keeps the revenue curve above the chord from $(x_i,r_i)$ to $(1,0)$. Put
\[
t:=\frac{r_i}{r_i+(1-x_i)p}.
\]
If $x_i=1$, then $D_i(p)=1=t$. Otherwise $p\le r_i/x_i$ gives $x_i\le t<1$, and concavity with $R_i(1)\ge0$ yields
\[
R_i(t)\ge\frac{1-t}{1-x_i}r_i=pt.
\]
The weak-tail quantile convention therefore gives $D_i(p)\ge t$, including at atoms. Since $r_i/p-x_i\ge0$,
\[
d_i(p)\le r_i-p t
=r_i\frac{r_i/p-x_i}{1+r_i/p-x_i}
\le\frac{r_i^2}{p}.
\]
\end{proof}

\subsection{Price assignments and inventory risk}\label{app:assignments}

Use the budget shares $\lambda_{i\ell}$, planned revenue $b$, inventory requirement $N\le K$, and aggregate shortfall $\mathcal E$ from Section~\ref{sec:realization}. All sums over $(i,\ell)$ below are restricted to pairs with $\lambda_{i\ell}>0$, so $r_i>0$ and $0<p_\ell\le w_i$ with $p_\ell\ge a$.

\paragraph{Actual assignments and the comparison payments.}
For each buyer, assign price $p_\ell$ with probability $\lambda_{i\ell}$ and no offer with the remaining probability. Draw one categorical lottery per original buyer, independently across buyers and independently of values. Approach buyers in decreasing assigned-price order and stop after $k$ sales.

Define buyer $i$'s potential payment by $W_i=p_\ell$ if outcome $\ell$ is assigned and $v_i\ge p_\ell$, and $W_i=0$ otherwise, including when no offer is assigned. To cap her expected payment conditional on the assigned price at $r_i$, keep this potential payment in the comparison with probability
\[
\tau_{i\ell}:=\min\left\{1,\frac{r_i}{p_\ell D_i(p_\ell)}\right\}.
\]
The denominator is positive because $D_i(p_\ell)\ge x_i>0$ on selected portions. Use an independent uniform draw for each buyer, independent of values and assignments. This gives independent hypothetical payments $\widetilde W_i$ satisfying $0\le\widetilde W_i\le W_i$. Expectations include values, price assignments, and, for $\widetilde W$, these additional draws. By construction,
\begin{equation}\label{eq:joint-demand}
\begin{aligned}
\sum_{i=1}^n\Pr_{\widetilde W}(\widetilde W_i>0)
&=\sum_{i,\ell}\lambda_{i\ell}\min\{D_i(p_\ell),r_i/p_\ell\}\le N\le k,\\
\sum_{i=1}^n\E_{\widetilde W}[\widetilde W_i]
&=\sum_{i,\ell}\lambda_{i\ell}\min\{p_\ell D_i(p_\ell),r_i\}
=b-\mathcal E.
\end{aligned}
\end{equation}

\paragraph{Revenue under the stock constraint.}

\begin{lemma}[Correlation gap]\label{lem:cg}
Let $W=(W_1,\ldots,W_n)$ have independent nonnegative coordinates, with $\sum_{i=1}^n\Pr_W(W_i>0)\le k$. Then
\[
\E_W[S_k(W)]\ge c_k\sum_{i=1}^n\E_W[W_i],
\qquad c_k=1-\frac{k^k}{e^k k!}.
\]
\end{lemma}

\begin{proof}
First consider a sum $X$ of independent Bernoulli variables with mean $\mu\le k$. We show that $\E[\min\{k,X\}]\ge c_k\mu$. Put $f(c):=\min\{k,c\}$. For nonnegative integers $c,j$,
\[
f(c+j)\le f(c)+j\bigl(f(c+1)-f(c)\bigr).
\]
Conditional on the other summands having total $c$, this inequality shows that replacing a Bernoulli variable of mean $p$ by an independent Poisson variable of mean $p$ cannot increase the expected truncated count. Replacing all summands gives
\[
\E[\min\{k,X\}]\ge g_k(\mu),
\qquad g_k(t):=\E[\min\{k,P_t\}],\quad P_t\sim\operatorname{Poi}(t).
\]
Differentiating the Poisson series gives
\[
g_k'(t)=\Pr(P_t<k),\qquad
g_k''(t)=-e^{-t}\frac{t^{k-1}}{(k-1)!}\le0.
\]
Thus $g_k$ is concave and $g_k(0)=0$, so $g_k(\mu)\ge(\mu/k)g_k(k)$. The Poisson recursion $j\Pr(P_k=j)=k\Pr(P_k=j-1)$ gives
\[
g_k(k)=k\bigl(1-\Pr(P_k=k)\bigr)=kc_k,
\]
proving the scalar comparison.

Now use the identity
\[
S_k(w)=\int_0^\infty\min\left\{k,\sum_{i=1}^n\ind{w_i>u}\right\}\,du
\]
for a deterministic nonnegative vector $w$. At each threshold $u\ge0$, the count $X_u:=\sum_i\ind{W_i>u}$ is a sum of independent Bernoulli variables with mean at most $\sum_i\Pr(W_i>0)\le k$. Tonelli's theorem and the scalar comparison yield
\[
\E_W[S_k(W)]\ge c_k\int_0^\infty\E[X_u]\,du
=c_k\sum_{i=1}^n\E_W[W_i].
\]
\end{proof}

\begin{proof}[Proof of Proposition~\ref{prop:reduction}(ii)]
The comparison payments are independent across original buyers. Equations~\eqref{eq:joint-demand} and~\eqref{eq:aggregate-shortfall} and Lemma~\ref{lem:cg} imply
\[
\begin{aligned}
\E_W[S_k(W)]&\ge\E_{\widetilde W}[S_k(\widetilde W)]\\
&\ge c_k(b-\mathcal E)\\
&\ge c_k\left(b-\sqrt{\frac{\eps\Rstar}{s}bN}\right).
\end{aligned}
\]
Decreasing-price sales collect $S_k(W)$ for every realization. Every assignment uses at most the plan's $L$ prices, so some fixed assignment has expected revenue over values at least the lottery average. Fix it before values are drawn to obtain the required deterministic mechanism.
\end{proof}

\subsection{Constants in the finite-market bound}\label{app:finite-bound}

We verify the choice of $s$ used in the proof of Proposition~\ref{prop:finite-bound} in Section~\ref{sec:assembly}. For $0<\eps\le\eps_0(k)$, set $s=(k\eps)^{1/3}$ and $\delta=s/(1-s)$. Then $s<1/2$, $C_L^\star\le2$, and
\[
\delta+\sqrt{\frac{C_L^\star k\eps}{s}}
\le\left(2+\sqrt2\right)s\le\frac12.
\]
Thus the denominator in that proof is positive. Using $1/(1-t)\le1+2t$ for $0\le t\le1/2$ and $C_L^\star/c_k\le4$ gives
\[
\frac{C_L^\star}{c_k}
\frac1{1-\delta-\sqrt{C_L^\star k\eps/s}}
\le\frac{C_L^\star}{c_k}+8(2+\sqrt2)(k\eps)^{1/3}.
\]

\subsection{Auxiliary inventory requirement and actual demand}\label{app:revenue-ceiling}

The following example illustrates why the implementation proof compares payments rather than identifying auxiliary inventory requirement with actual demand.

\begin{figure}[t]
\centering
\begin{tikzpicture}
\begin{axis}[
    width=0.88\linewidth, height=6.1cm,
    xmin=0, xmax=0.012, ymin=0, ymax=0.011,
    xlabel={Quantity $q$}, ylabel={Revenue}, scaled ticks=false,
    xtick={0,0.004,0.008,0.012}, ytick={0,0.004,0.008},
    tick label style={font=\scriptsize,/pgf/number format/fixed,/pgf/number format/precision=3},
    label style={font=\small}, grid=major, grid style={gray!15},
    legend pos=north west,
    legend style={font=\small,fill=white,draw=none,cells={anchor=west}},
]
\addplot[blue!75!black,very thick]
    coordinates {(0,0) (0.001,0.0018) (0.012,0.01049)};
\addlegendentry{$R(q)$}
\addplot[red!75!black,dashed,thick]
    coordinates {(0,0) (0.012,0.0108)};
\addlegendentry{$pq$}
\draw[densely dotted,gray] (axis cs:0,0.0018) -- (axis cs:0.002,0.0018);
\draw[densely dotted,gray] (axis cs:0.001,0) -- (axis cs:0.001,0.0018);
\draw[densely dotted,gray] (axis cs:0.002,0) -- (axis cs:0.002,0.0018);
\draw[densely dotted,gray] (axis cs:0.009181818,0) -- (axis cs:0.009181818,0.008263636);
\addplot[only marks,mark=*,mark size=2pt,blue!75!black,forget plot]
    coordinates {(0.001,0.0018) (0.009181818,0.008263636)};
\addplot[only marks,mark=o,mark size=2.5pt,red!75!black,mark options={fill=white},forget plot]
    coordinates {(0.002,0.0018)};
\node[anchor=west,font=\scriptsize] (reference) at (axis cs:0.0022,0.0042) {Reference $(x,r)$};
\draw[->,gray!80] (reference.west) -- (axis cs:0.001,0.0018);
\node[anchor=west,font=\scriptsize] (quantity) at (axis cs:0.0033,0.0013) {Auxiliary inventory requirement $r/p$};
\draw[->,gray!80] (quantity.west) -- (axis cs:0.002,0.0018);
\node[anchor=east,font=\scriptsize] at (axis cs:0.0088,0.0095) {Actual demand $D(p)$};
\end{axis}
\end{tikzpicture}
\caption{For the buyer in Example~\ref{ex:demand-bill}, the reference point $(x,r)$ determines the auxiliary budget. At price $p$, the open circle on the price ray $pq$ has revenue $r$ and inventory requirement $r/p$. Actual demand is the farther intersection of the ray with $R(q)$, where revenue exceeds $r$. Thus the auxiliary inventory requirement need not bound actual acceptance probability.}
\label{fig:revenue-demand-bill}
\end{figure}
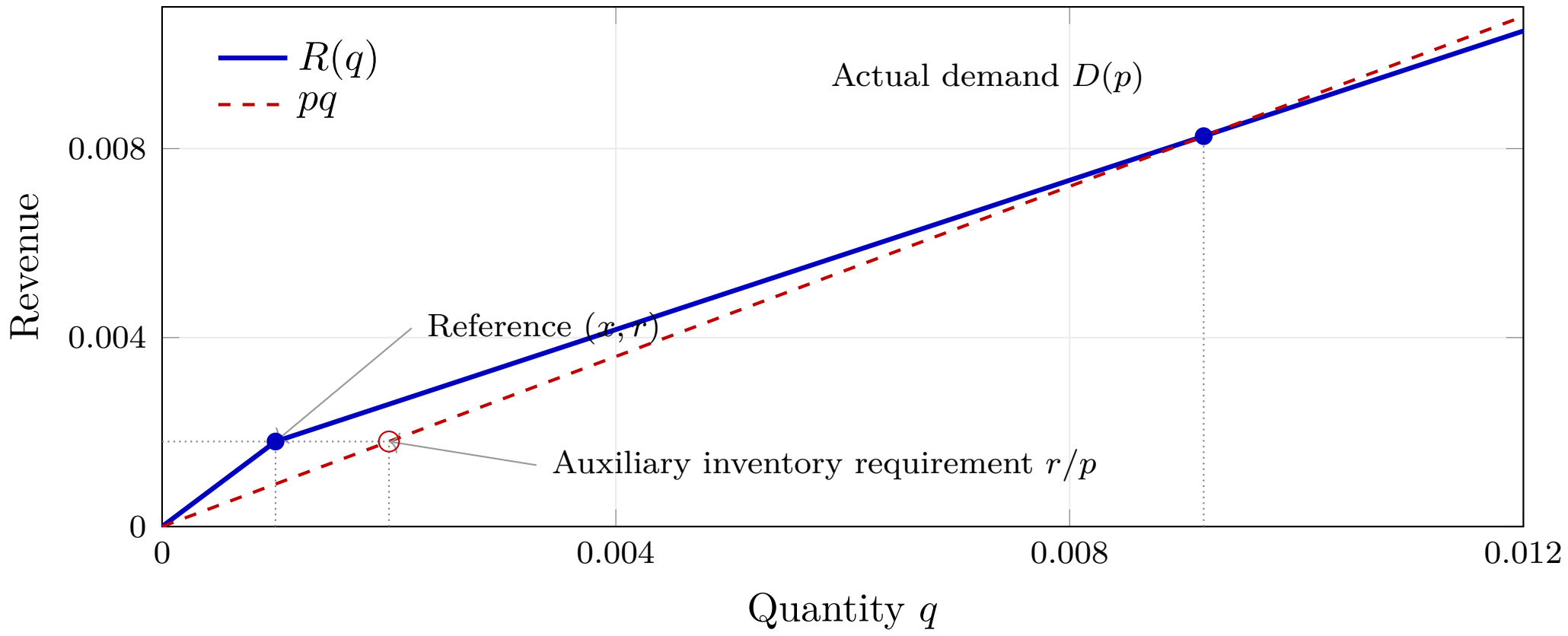

\begin{example}[Auxiliary inventory requirement and actual demand]\label{ex:demand-bill}
Consider one buyer with revenue curve $R(q)=\min\{1.8q,\,0.79q+0.00101\}$ and threshold $a=0.8$. The maximizer of $R(q)-aq$ is $x=0.001$, with revenue $r=0.0018$ and reference price $w=1.8$. Her maximum excess revenue is $0.001$, so the ceiling holds with $\eps\Rstar/s=0.001$. At price $p=0.9$,
\[
\frac rp=0.002,
\qquad D(p)=\frac{0.00101}{0.9-0.79}\approx0.00918.
\]
Actual demand is more than four times the auxiliary inventory requirement, even though $pD(p)>r$. Figure~\ref{fig:revenue-demand-bill} shows the distinction. This is an individual revenue-curve example, not a complete market.
\end{example}

\section{Markets Approaching the Sharp Bounds}\label{app:sharpness-transfer}

This appendix proves the revenue comparisons used in Section~\ref{sec:coarseness-constants}.

Fix an integer $T\ge1$ of premium types, a baseline price $a>0$, and capacity $K>0$. Type $t\in\{1,\ldots,T\}$ has budget $r_t>0$ and reference price $w_t>a$, with types ordered by decreasing $w_t$. Total premium spending is $\Pi$, and collecting it at individual values requires $Q$ units. Filling the remaining capacity at the baseline price gives unrestricted revenue $B$:
\[
\Pi:=\sum_{t=1}^T r_t,\qquad Q:=\sum_{t=1}^T\frac{r_t}{w_t}\le K,
\qquad B:=\Pi+a(K-Q).
\]
Define the unit-cost profile $u$ on $(0,B]$ to equal $1/w_t$ on the interval of budget $r_t$, followed by $1/a$ on the baseline spending interval of length $a(K-Q)$. This is an auxiliary market with budgets $r_t$ at values $w_t$ and a final budget $a(K-Q)$ at value $a$, omitted when zero. Its unrestricted revenue is $B$, and its $L$-price revenue is $V_L(u,K)$. These are fixed auxiliary data and need not arise from the ex-ante construction.

\begin{proposition}[From auxiliary markets to large markets]\label{prop:sharpness-transfer}
For every integer stock $k\ge1$ and all sufficiently large integers $m$, there is a market $F_{k,m}$ with independent regular values and finite means that is $\eps_m$-large for
\[
\eps_m:=\frac{\max_{1\le t\le T}r_t}{Ka m}.
\]
Writing $w_{\max}:=\max_{1\le t\le T}w_t$, its revenues satisfy
\[
\begin{aligned}
\frac{K}{k}\LPP_{\infty}(F_{k,m})
&\ge B-\frac{Kw_{\max}}{\sqrt{k}},\\
\frac{K}{k}\LPP_L(F_{k,m})&\le V_L(u,K)
\qquad\text{for every }L\ge1.
\end{aligned}
\]
\end{proposition}

\subsection{Constructing the buyers}

For each premium type $t$, create $m$ buyers with independent values and demand curve
\[
D_{t,m}(p)=
\begin{cases}
1,&0\le p<c_t,\\
c_t/p,&c_t\le p\le w_t,\\
0,&p>w_t,
\end{cases}
\qquad c_t:=\frac{kr_t}{Km}.
\]
Choose $m$ large enough that every $c_t<a$, and add $k+1$ deterministic buyers of value $a$. Each premium distribution has bounded support $[c_t,w_t]$ and the concave revenue curve
\[
R_{t,m}(q)=\min\{w_tq,c_t\},\qquad q\in[0,1].
\]
Thus it is regular with finite positive mean. Its nonnegative virtual value is $w_t$ with probability $c_t/w_t$ and zero otherwise. Write $F_{k,m}$ for the resulting market and abbreviate its revenues by $\Rstar$, $\LPP_{\infty}$, and $\LPP_L$.

Removing a premium buyer reduces optimal revenue by at most $c_t$, by the pivotal identity. Removing any baseline buyer changes nothing because at least $k$ others of value $a$ remain. Since $\Rstar\ge ak$, the market is $\eps_m$-large with $\eps_m$ as in Proposition~\ref{prop:sharpness-transfer}.

\subsection{Comparing unrestricted and coarse prices}

Offer each type its reference price $w_t$, in decreasing order, and then fill any remaining stock at $a$. Let $X$ be the number of the $Tm$ premium buyers whose value equals their type-specific top atom $w_t$. Its mean is $(k/K)Q\le k$, and its variance is at most $k$. Valuing every premium acceptance at its markup above $a$ and adding the baseline $ak$ gives ideal expected revenue $(k/K)B$. Enforcing the stock limit reduces this revenue by at most $w_{\max}(X-k)_+$, so
\begin{equation*}
\begin{aligned}
\Rstar\ge\LPP_{\infty}
&\ge\frac{k}{K}B-w_{\max}\E_X[(X-k)_+]\\
&\ge\frac{k}{K}B-w_{\max}\sqrt{k}.
\end{aligned}
\end{equation*}
The last step uses $\E_X[X]\le k$ and Cauchy--Schwarz.

To bound every $L$-price mechanism, fix any assignment and list its distinct finite prices above $a$ as $p_1>\cdots>p_J>a$, where $0\le J\le L$. For $t=1,\ldots,T$ and $j=1,\ldots,J$, let $z_{tj}$ be $K/k$ times the expected payment collected from type-$t$ buyers at $p_j$, averaging over values and the resulting stock truncation. Since every buyer receives at most one price and $pD_{t,m}(p)\le c_t$ for every finite $p\ge0$,
\begin{equation*}
\sum_{j=1}^J z_{tj}\le r_t,
\qquad z_{tj}=0\ \text{if }p_j>w_t.
\end{equation*}
In auxiliary units, total premium revenue is $z:=\sum_{t=1}^T\sum_{j=1}^J z_{tj}$ and its expected inventory use is $N_0:=\sum_{t=1}^T\sum_{j=1}^J z_{tj}/p_j$, with both sums zero when $J=0$. Thus $N_0\le K$. Payments collected at prices at most $a$ are at most $(k/K)a(K-N_0)$.

Apply the rearrangement in the proof of Lemma~\ref{lem:auxiliary-plans} to these collected payments: moving each type's spending to its highest eligible price, selecting the first $z$ budget dollars, and raising each block to its final reference price gives a premium prefix with revenue $z$, inventory requirement $\widetilde N\le N_0$, and at most $J$ blocks.

If the assignment uses no price at most $a$, this prefix has at most $L$ blocks and contains all its revenue, giving revenue at most $(k/K)V_L(u,K)$. If a price at most $a$ is used, the prefix has at most $L-1$ blocks. Value the remaining capacity $K-\widetilde N$ at $a$ and append the revenue endpoint
\[
b:=z+a(K-\widetilde N).
\]
Since every remaining premium dollar costs at most $1/a$ units, $\int_z^\Pi u(R)\,dR\le(\Pi-z)/a$. Hence this endpoint lies in the full profile:
\[
b\le z+a\left(K-\int_0^z u(R)\,dR\right)
\le\Pi+a(K-Q)=B.
\]
The added block costs at most $(b-z)/a=K-\widetilde N$, since $u\le1/a$. It is therefore a feasible plan with at most $L$ blocks, and
\[
z+a(K-N_0)\le b\le V_L(u,K).
\]
Both cases give $\LPP_L\le(k/K)V_L(u,K)$, proving Proposition~\ref{prop:sharpness-transfer}.

\section{Quasi-Regular Distributions}\label{app:quasireg}

A distribution is \emph{quasi-regular} if $R(q)/(1-q)$ is nondecreasing on $[0,1)$. Equivalently, its revenue curve stays above the line from any point $(q,R(q))$ to $(1,0)$ at larger quantiles. Concavity implies this condition, which also permits marginal revenue to rise. This enlarges the permitted demand distributions: every order statistic of independent, possibly differently distributed regular draws is quasi-regular \citep[Theorem~3]{FengJin25}. Thus the model allows values generated as maxima or other order statistics, subject to its independence and finite-mean assumptions.

Write $\cF_k^{\mathrm{qr}}(\eps)$ for the class $\cF_k(\eps)$ with regularity replaced by quasi-regularity, keeping the finite-mean and positive-revenue requirements.

\begin{proposition}[Quasi-regular extension]
Proposition~\ref{prop:finite-bound} and Theorem~\ref{thm:sharp-market} hold with $\cF_k(\eps)$ replaced by $\cF_k^{\mathrm{qr}}(\eps)$, with the same constants.
\end{proposition}

\begin{proof}
Fix $F\in\cF_k^{\mathrm{qr}}(\eps)$ and $0<s<1/2$, and abbreviate $\Rstar=\Rstar(F)$. Let $\bar R_i$ be the least concave majorant of $R_i$, and let $\bar\varphi_i$ be its nonincreasing marginal revenue. The weak-tail revenue curve is upper semicontinuous, so its majorant is affine between contacts and every point is a mixture of at most two contacts with $R_i$. Finite means give $\bar R_i(0)=0$ and continuity at zero: a two-point representation at $q$ bounds its two revenue contributions by $\sup_{t\le q}R_i(t)$ and $R_i(q)$, both vanishing as $q\downarrow0$. Myerson's ironing identity gives $\Rstar=\E[S_k(Y)]$ with $Y_i=(\bar\varphi_i(U_i))_+$, also in every submarket \citep{Myerson81}.

Choose the threshold $a$ and ex-ante allocation $x$ as in Section~\ref{sec:benchmark}, using $\bar R_i$ and its marginal revenues, with the same quantity tie-break. Put $K=\sum_i x_i$ and $B=\sum_i\bar R_i(x_i)$. The ex-ante relaxation is valid because an incentive-compatible interim allocation is a mixture of threshold offers, and concavity of $\bar R_i$ bounds their average revenue at the average service probability. The threshold, allocation, and pivotal arguments in Appendix~\ref{app:toolkit} therefore give
\[
\Rstar\le B+a(k-K)\le B+\frac{s}{1-s}\Rstar,
\qquad
\max_{q\in[0,1]}\{\bar R_i(q)-aq\}\le\frac{\eps}{s}\Rstar.
\]

The additional step is to express the selected ironed revenue through actual price opportunities. Let $t_i$ be the rightmost maximizer of $\bar R_i(q)-aq$. Optimality gives $x_i\le t_i$. Use $x_i$ itself if it is a contact; otherwise use the endpoints of its affine face, whose slope is at least $a$ and whose endpoints lie in $[0,t_i]$. For $\bar R_i(q)-aq$, a rising face places the rightmost maximizer at or beyond its right endpoint, while a flat face consists of maximizers. After omitting a zero endpoint, there are at most two positive contact quantiles $q_{ij}$ with revenues $r_{ij}:=R_i(q_{ij})=\bar R_i(q_{ij})$ and weights $\omega_{ij}\ge0$ such that
\[
x_i=\sum_j\omega_{ij}q_{ij},\qquad
\bar R_i(x_i)=\sum_j\omega_{ij}r_{ij},\qquad
\sum_j\omega_{ij}\le1.
\]
These sums are empty when $x_i=0$. Concavity gives $r_{ij}\ge aq_{ij}$ at every chosen contact. Each positive weighted revenue $\omega_{ij}r_{ij}$ becomes an auxiliary budget at reference price $w_{ij}:=r_{ij}/q_{ij}$. Omitting zero budgets, their total revenue is $B$ and their inventory requirement at these prices is at most $K$.

Quasi-regularity supplies the shortfall bound at each contact. For $0<p\le w_{ij}$, put $t=r_{ij}/[r_{ij}+(1-q_{ij})p]$. If $q_{ij}<1$, then $q_{ij}\le t<1$ and the defining condition gives $R_i(t)\ge(1-t)r_{ij}/(1-q_{ij})=pt$, hence $D_i(p)\ge t$. If $q_{ij}=1$, the same conclusion follows from $D_i(p)=1=t$. For $p\ge a$, this demand bound and the ironed revenue ceiling yield, by the algebra in Appendix~\ref{app:shortfalls},
\[
d_{ij}(p):=(r_{ij}-pD_i(p))_+
\le\min\left\{\frac{\eps}{s}\Rstar,\frac{r_{ij}^2}{p}\right\}.
\]

Fix a feasible auxiliary plan. Let $\lambda_{ij\ell}r_{ij}$ be the revenue selected from contact $j$ of buyer $i$ in price block $\ell$. Thus $\sum_\ell\lambda_{ij\ell}\le\omega_{ij}$. Sums below run over selected portions with positive revenue. The plan's revenue, inventory requirement, and aggregate shortfall satisfy
\[
\begin{aligned}
b&=\sum_{i,j,\ell}\lambda_{ij\ell}r_{ij},
\qquad N=\sum_{i,j,\ell}\lambda_{ij\ell}\frac{r_{ij}}{p_\ell}\le K,\\
\mathcal E&:=\sum_{i,j,\ell}\lambda_{ij\ell}d_{ij}(p_\ell)
\le\sqrt{\frac{\eps\Rstar}{s}bN},
\end{aligned}
\]
where the last inequality is the weighted Cauchy--Schwarz calculation in~\eqref{eq:aggregate-shortfall}.

For each original buyer, draw one offer lottery whose outcome $(j,\ell)$ assigns price $p_\ell$ with probability $\lambda_{ij\ell}$. The remaining probability gives no offer; it is nonnegative because $\sum_{j,\ell}\lambda_{ij\ell}\le1$. Draw these lotteries independently across buyers and independently of values. Conditional on outcome $(j,\ell)$, keep an accepted payment in the hypothetical comparison with probability $\min\{1,r_{ij}/[p_\ell D_i(p_\ell)]\}$, using independent draws. As in Appendix~\ref{app:assignments}, comparison payments are bounded by actual potential payments, have total expected demand at most $N$, and have total expected payment $b-\mathcal E$. They are independent across original buyers. Lemma~\ref{lem:cg} and a fixed assignment attaining at least the lottery average therefore give a deterministic $L$-level mechanism with expected revenue at least
\[
c_k\left(b-\sqrt{\frac{\eps\Rstar}{s}bN}\right).
\]
Thus Proposition~\ref{prop:reduction} holds for the enlarged class, and Section~\ref{sec:assembly} gives the same finite-market bound. The sharpness markets in Section~\ref{sec:coarseness-constants} are regular and hence quasi-regular, so the limit argument in Section~\ref{sec:joint-growth} also gives the same sharp ratios.
\end{proof}

\bibliographystyle{apalike}
\bibliography{reference}

\end{document}